\documentclass[AMA,STIX1COL,doublespace]{WileyNJD-v2}
\pdfoutput=1
\usepackage{microtype}
\usepackage[utf8]{inputenc}
\usepackage[T1]{fontenc}
\usepackage{dsfont}
\usepackage{physics}
\usepackage{nicefrac}
\usepackage{bm}
\usepackage{xcolor}
\usepackage{enumitem}
\usepackage{graphicx}
\usepackage{varioref}
\usepackage{tabularx}
\usepackage{booktabs}
\usepackage[caption=false]{subfig}
\usepackage[section]{placeins}
\usepackage{csquotes}
\usepackage{cleveref}
\usepackage[final,commandnameprefix=ifneeded]{changes}

\definechangesauthor[name={Philipp Diercks}, color=blue]{pd}

\DeclareMathOperator*{\dist}{dist}

\articletype{Article Type}%

\received{<day> <Month>, <year>}
\revised{<day> <Month>, <year>}
\accepted{<day> <Month>, <year>}

\newcommand{\BeamMvnRrfMin}{1.77}
\newcommand{\BeamMvnRrfMax}{19.76}
\newcommand{\BeamMvnRrfMean}{7.84}
\newcommand{\BeamMvnExtMin}{0.19}
\newcommand{\BeamMvnExtMax}{0.5}
\newcommand{\BeamMvnExtMean}{0.27}

\newcommand{\BeamNormalRrfMin}{1.64}
\newcommand{\BeamNormalRrfMax}{5.66}
\newcommand{\BeamNormalRrfMean}{3.49}
\newcommand{\BeamNormalExtMin}{0.18}
\newcommand{\BeamNormalExtMax}{0.6}
\newcommand{\BeamNormalExtMean}{0.24}

\newcommand{\BeamFomNdofs}{4204602.0}
\newcommand{\BeamFomAssembly}{2.92}
\newcommand{\BeamFomSolve}{35.79}

\newcommand{\BeamMvnRomNdofs}{3923.8}
\newcommand{\BeamMvnRomMaxModes}{15.2}
\newcommand{\BeamMvnRomAssembly}{0.3}
\newcommand{\BeamMvnRomSolve}{0.11}

\newcommand{\BeamNormalRomNdofs}{3925.8}
\newcommand{\BeamNormalRomMaxModes}{12.1}
\newcommand{\BeamNormalRomAssembly}{0.29}
\newcommand{\BeamNormalRomSolve}{0.1}

\newcommand{\BlockMvnRrfMin}{0.61}
\newcommand{\BlockMvnRrfMax}{2.82}
\newcommand{\BlockMvnRrfMean}{1.24}
\newcommand{\BlockMvnExtMin}{0.05}
\newcommand{\BlockMvnExtMax}{0.38}
\newcommand{\BlockMvnExtMean}{0.1}

\newcommand{\BlockNormalRrfMin}{0.48}
\newcommand{\BlockNormalRrfMax}{1.32}
\newcommand{\BlockNormalRrfMean}{0.81}
\newcommand{\BlockNormalExtMin}{0.04}
\newcommand{\BlockNormalExtMax}{0.18}
\newcommand{\BlockNormalExtMean}{0.08}

\newcommand{\BlockFomNdofs}{76402.0}
\newcommand{\BlockFomAssembly}{0.05}
\newcommand{\BlockFomSolve}{0.53}

\newcommand{\BlockMvnRomNdofs}{492.0}
\newcommand{\BlockMvnRomMaxModes}{32.8}
\newcommand{\BlockMvnRomAssembly}{0.0}
\newcommand{\BlockMvnRomSolve}{0.0}

\newcommand{\BlockNormalRomNdofs}{492.0}
\newcommand{\BlockNormalRomMaxModes}{29.4}
\newcommand{\BlockNormalRomAssembly}{0.0}
\newcommand{\BlockNormalRomSolve}{0.0}

\newcommand{\LpanelMvnRrfMin}{2.17}
\newcommand{\LpanelMvnRrfMax}{38.61}
\newcommand{\LpanelMvnRrfMean}{20.04}
\newcommand{\LpanelMvnExtMin}{0.26}
\newcommand{\LpanelMvnExtMax}{0.99}
\newcommand{\LpanelMvnExtMean}{0.55}

\newcommand{\LpanelNormalRrfMin}{1.79}
\newcommand{\LpanelNormalRrfMax}{6.54}
\newcommand{\LpanelNormalRrfMean}{4.58}
\newcommand{\LpanelNormalExtMin}{0.23}
\newcommand{\LpanelNormalExtMax}{0.84}
\newcommand{\LpanelNormalExtMean}{0.39}

\newcommand{\LpanelFomNdofs}{5042402.0}
\newcommand{\LpanelFomAssembly}{3.96}
\newcommand{\LpanelFomSolve}{53.98}

\newcommand{\LpanelMvnRomNdofs}{4522.0}
\newcommand{\LpanelMvnRomMaxModes}{34.5}
\newcommand{\LpanelMvnRomAssembly}{0.26}
\newcommand{\LpanelMvnRomSolve}{0.2}

\newcommand{\LpanelNormalRomNdofs}{4522.0}
\newcommand{\LpanelNormalRomMaxModes}{22.2}
\newcommand{\LpanelNormalRomAssembly}{0.26}
\newcommand{\LpanelNormalRomSolve}{0.19}

\newcommand{\LpanelFieldsNumModes}{12}
\newcommand{\NumReal}{20}

\begin{document}

\title{Multiscale modeling of linear elastic heterogeneous structures via localized model order reduction}

\author[1]{Philipp Diercks*}
\author[tue]{Karen Veroy}
\author[1]{Annika Robens-Radermacher}
\author[1]{Jörg F. Unger}

\authormark{\textsc{Diercks et al}}

\address[1]{\orgdiv{Department 7.7 Modeling and Simulation}, \orgname{Bundesanstalt für Materialforschung und -prüfung (BAM)}, \orgaddress{Unter den Eichen 87, 12205 \state{Berlin}, \country{Germany}}}

\address[tue]{\orgdiv{Centre for Analysis, Scientific Computing and Applications (CASA) Department of Mathematics and Computer Science}, \orgname{University of Eindhoven}, \orgaddress{P.O. Box 513, 5600 MB \state{Eindhoven}, \country{The Netherlands}}}

\corres{*Philipp Diercks, \orgaddress{Unter den Eichen 87, 12205 \state{Berlin}, \country{Germany}}, \email{philipp.diercks@bam.de}}

\presentaddress{\orgaddress{Unter den Eichen 87, 12205 \state{Berlin}, \country{Germany}}}

\abstract[Abstract]{%
In this paper, a methodology for fine scale modeling of large scale
linear elastic structures is proposed, which combines the variational multiscale method,
domain decomposition and model order reduction.
The influence of the fine scale on the coarse scale is modelled by the use
of an additive split of the displacement field, addressing applications
without a clear scale separation.
Local reduced spaces are constructed by solving an oversampling problem with 
random boundary conditions.
Herein, we inform the boundary conditions by a global reduced problem and compare
our approach using physically meaningful correlated samples with existing
approaches using uncorrelated samples.
The local spaces are designed such that the local contribution of each subdomain
can be coupled in a conforming way, which also preserves the sparsity
pattern of standard finite element assembly procedures.
Several numerical experiments show the accuracy and efficiency of the method, as well 
as its potential to reduce the size of the local spaces and the number
of training samples compared to the uncorrelated sampling.
}


\jnlcitation{\cname{%
\author{P. Diercks}, 
\author{K. Veroy}, 
\author{A. Robens-Radermacher}, and 
\author{J. F. Unger}} (\cyear{2023}), 
\ctitle{Multiscale modeling of linear elastic heterogeneous structures via localized model order reduction}, \cjournal{International Journal for Numerical Methods in Engineering}, \cvol{2023;XX:X-X}.}

\maketitle

\section{Introduction}%
\label{sec:introduction}

\subsection{Multiscale modeling and model order reduction}%
\label{sub:multiscale_modeling_and_model_order_reduction}

Many problems in science and engineering involve multiple scales. 
With large heterogeneities present in spatial scales,
it is often insufficient to assume a homogeneous material in the analysis of a mechanical structure.
For example, the dispersed phases (particles or fibers) in a composite material may lead to fluctuations in the displacement field which cannot be captured by the phenomenological macroscale model. 
Therefore, in analyzing large scale structures, it is necessary to take into account the materials' fine scale heterogeneity to more accurately model the structure's behaviour.

It is often sufficient to predict macroscopic properties of the multiscale system based on a representative volume element (RVE) that preserves the geometrical complexity of the heterogeneous microstructure and accurately predicts effective material parameters.
Approaches based on computational homogenization, such as the FE$^2$ method (see e.\,g.~\cite{MieheKoch2002,FeyelCaboche2000,GeersKouznetsovaEtAl2010a,GeersKouznetsovaEtAl2010}) mitigate the issue of computational cost compared to full fine scale simulations, but the nested solution procedure is still a demanding task.
Therefore, many approaches~\cite{YvonnetHe2007,GouryKerfridenEtAl2014,HernandezOliverEtAl2014,GuoRokosVeroy2021} combining the FE$^2$ approach with model reduction of the fine scale problem exist.
These approaches rest on the assumption of \textit{separation of scales} and the
existence of an RVE\@; however, this is not the case in many applications,
e.\,g.\, for composite structures where simply the dimension of the fine scale
features is not much smaller than the macroscopic dimension, therefore breaking
the MMM-principle defined in~\cite{Hashin1983}, or in the presence of macroscopic
cracks emerging from the localization of microdefects~\cite{Gitman2006,GitmanEtAl2007}.

Thus, methods which address both scales simultaneously are needed.
Standard multiscale methods have emerged from the variational approaches to numerical homogenization, such as the variational multiscale method (VMM)~\cite{Hughes1995,HughesFeijooEtAl1998} or the multiscale finite element method (MsFEM)~\cite{HW1997}, which aim for a correction or stabilization of the conventional (coarse grid) discretization by including (unresolved) fine scale information into the global problem.
Important developments of the VMM include the works by M.\,Larson and A.\,Målqvist~\cite{LarsonMalqvist2005,LarsonMalqvist2007,LarsonMalqvist2009} and the local orthogonal decomposition (LOD)~\cite{MalqvistPeterseim2014}.
For a more detailed discussion on the history of numerical homogenization in the absence of a clear separation of scales we refer to Altmann~et~al.~\cite{AltmannHenningPeterseim2021}.

In addition, the significant increase in computational cost entailed with the resolution of the fine scale features in the numerical model makes the direct
solution of the problem infeasible.
In this work, this is addressed by the use of model order reduction techniques (see the textbooks~\cite{HesthavenRozzaStamm2016,QuarteroniManzoniNegri2016} for an introduction to the topic).
The high dimensional numerical problem (also termed high-fidelity approximation or full order model (FOM)) is replaced by a reduced order model (ROM) of small dimension, which is achieved by the projection of the original system of equations upon a low-dimensional subspace of the high-dimensional space in which the solution lives.
A key point is the construction of the reduced basis, which spans the low dimensional subspace, from a set of suitably selected high-fidelity solutions. 
In (now standard) reduced basis (RB) methods, the so-called snapshots are selected via the weak greedy algorithm~\cite{PrudhommeEtAl2002,VeroyEtAl2003}.
Another popular method for subspace construction is the proper orthogonal decomposition (POD)~\cite{HolmesEtAl2012,KunischVolkwein2002}.
However, efficient reduction of nonlinear problems in mechanics still poses a challenge due to the repeated evaluation of the nonlinear operator over the full domain.
Among others, well-known techniques to address this issue are the empirical interpolation method (EIM)~\cite{BarraultEtAl2004} and its discrete variant~\cite{Chaturantabut2010}, the hyper-reduction~\cite{Ryckelynck2005,Ryckelynck2009}, the energy-conserving sampling and weighting method~\cite{FarhatEtAl2014,FarhatEtAl2015} or the empirical cubature method~\cite{HernandezCaicedoFerrer2016}.
Moreover, more recent approaches~\cite{GuoHesthaven2018,RaissiEtAl2019} make use of machine learning methods to construct ROMs for nonlinear problems.

In the case of full fine scale simulations, limitations of established model order reduction techniques become apparent;
examples of such limitations include prohibitively large reduced spaces due to high dimensional parameter spaces or computationally expensive offline phases due to large computational domains.
To alleviate these shortcomings, methods combining multiscale methods, domain decomposition and model order reduction were developed.
Approaches of this kind are known as \textit{localized model order reduction methods}, and an extensive review is given by Buhr~et~al.~\cite{BuhrEtAlInBook2020}.
The main idea is the construction of local reduced spaces on subdomains, i.\,e.\ parts of the global domain, which are then coupled (either in a conforming or non-conforming way) to obtain a global approximation. 

\subsection{Contributions and relation to previous work}%
\label{sub:contributions_and_relation_to_previous_work}

In this work, we aim to provide a computationally efficient framework
for multiscale modeling of linear heterogeneous structures that is able to incorporate
localization phenomena as described in
\cref{sub:multiscale_modeling_and_model_order_reduction}.
While this contribution is limited to the linear case,
we suggest an approach that addresses both scales simultaneously and flexibly,
with a view towards future extensions to nonlinear material behaviour.

The proposed methodology features an additive split of the displacement field
into coarse and fine scale parts, based on the VMM\@.
The coarse scale basis functions are computed directly by extending standard
finite element shape functions on the boundary of local subdomains into
the interior of the respective subdomains.
Local approximation spaces for the fine scale part are constructed by exploiting possible fine scale solutions for a coarse grid element using the concept of oversampling first introduced in the context of the MsFEM~\cite{HW1997}.
To this end, a so-called transfer eigenvalue problem~\cite{BabuskaLipton2011,SmetanaPatera2016} (or oversampling problem) yielding local reduced spaces which are optimal in the sense of Kolmogorov is solved and the associated transfer operator is approximated by random sampling~\cite{BS2018}.
The novelty consists in the use of a \textit{multivariate normal} distribution \added[id=pd]{with non-zero mean given by the solution of a reduced global problem, and a covariance matrix with squared exponential kernel}
to sample the random boundary conditions.
To this end, algorithm 1 of~\cite{BS2018} is modified to inform the boundary
conditions of the oversampling problem by the solution of a reduced global problem,
incorporating the macroscopic displacement state of the structure of interest
into the training data.
Note that a similar approach to build local reduced spaces using interface basis functions (Lagrangian or Fourier bases) as boundary conditions was proposed in Iapichino~et~al.~\cite{IapichinoQuarteroniRozza2016}.
The difference is that Iapichino~et~al.\ prescribed these interface basis functions on the boundary of the subdomain of interest directly.
In our approach, the macroscopic displacement state of the global structure of interest is prescribed on the boundary of the oversampling domain, making it more suitable for the construction of reduced spaces tailored to the solution of the partial differential equation~(PDE) in that area of the domain.
Moreover, we further restrict the fine scale solutions obtained from the oversampling problem to the edges of the target subdomain and \deleted{by means of POD,} construct a reduced fine scale edge basis, separately for each edge in the partition of the global domain.
Then, the fine scale edge basis functions are again extended into the interior
of the respective subdomains.
We note that such a procedure to construct a conforming, localized reduced order approximation is outlined in a more general form in the review by Buhr~et~al.~\cite{BuhrEtAlInBook2020}.
Finally, the fine scale subdomain basis functions are problem-dependent local functions which are continuous on subdomain boundaries and thus yield a conforming approximation.
This decomposition of the fine scale part in its respective edge parts is favorable since the resulting discrete equation system of the (global) ROM preserves the sparsity pattern and computational complexity of standard finite element methods.

In view of future extensions of the method to the nonlinear case, we expect that the impact of incorporating the local deformation state of the structure of interest in the construction of the local reduced basis will be more significant. Also, the amplitude of the boundary conditions prescribed in the oversampling problem is relevant in nonlinear problems which may be challenging in the case of random boundary conditions.
Due to the resemblance of the constructed empirical fine scale basis functions with hierarchical FE shape functions (see e.g.~\cite{ZienkiewiczTaylor2000}), it is then possible to incorporate strategies from the field of adaptive refinement ($p$-refinement).

As an alternative to the decomposition outlined above, the generalized finite element method~(GFEM)~\cite{BabuskaGalozOsborn1994,BabuskaMelenk1997,BabuskaBanerjeeOsborn2004} can be used to construct a global approximation from local reduced spaces.
To this end, local reduced basis functions are multiplied with standard finite element shape functions to create a partition of unity.

The remainder of this article is organised as follows.
First, in \cref{sec:problem_setting_and_modeling}, the problem setting and full order model are described. 
The proposed method is explained in \cref{sec:multiscale_method}, comprising the construction of local approximation spaces in~\cref{sub:construction_of_local_approximation_spaces} and the assembly of the reduced order model in~\cref{sub:reduced_order_model}.
Numerical examples illustrating the performance of the suggested approach are discussed in \cref{sec:numerical_experiments}.
Concluding remarks and an outlook are given in \cref{sec:conclusion}.

\section{Problem setting and modeling}%
\label{sec:problem_setting_and_modeling}
While the method could be applied to other linear PDEs, only the balance of linear momentum in the static case on a large computational domain $\varOmega_{\mathrm{gl}} \subset\mathbb{R}^d$ (the suffix `gl' stands for global) is considered, with boundary $\partial\varOmega_{\mathrm{gl}} = \varSigma_{\mathrm{N}}\cup\varSigma_{\mathrm{D}}$, where $\varSigma_{\mathrm{N}}$ and $\varSigma_{\mathrm{D}}$ denote Neumann and Dirichlet boundaries, respectively, and $d=2, 3$ is the spatial dimension.
Without loss of generality, volumetric forces are neglected, and the displacement solution $\bm{u}_{\mathrm{gl}}$ is sought such that
\begin{align}
    \label{eq:strong_bvp}
    \begin{split}
	    - \nabla \vdot \bm{\sigma} (\nabla\bm{u}_{\mathrm{gl}}) &= 0 \quad  \mathrm{in}\;\varOmega_{\mathrm{gl}}\,,\\
	    \bm{\sigma}(\nabla\bm{u}_{\mathrm{gl}}) \vdot \bm{n} &= \hat{\bm{t}} \quad \mathrm{on}\; \varSigma_{\mathrm{N}}\,,\\
	    \bm{u}_{\mathrm{gl}} &= \bm{g}_{\mathrm{D}} \quad  \mathrm{on}\;\varSigma_{\mathrm{D}}\,.
    \end{split}
\end{align}
Here the Cauchy stress tensor $\bm\sigma$ for one of the $M$ material components of the heterogeneous linear elastic material
is given by
\begin{equation}
	\label{eq:hooke}
	\bm{\sigma}_m = \lambda^1_m ({\varepsilon}(\bm{u}_{\mathrm{gl}}) \vdot\!\vdot\, \mathds{1}) \mathds{1} + 2 \lambda^2_m {\varepsilon}(\bm{u}_{\mathrm{gl}})\,, \quad \mathrm{with} \quad
	m=1, \ldots, M\,,
\end{equation}
where $\lambda_m^1$ and $\lambda_m^2$ are Lame's constants.
The linear strain operator is denoted by ${\varepsilon}(\bm{v}) = \frac{1}{2}\left(\nabla\bm{v} + {\nabla\bm{v}}^{\mathrm{T}}\right)$.
Moreover, $\bm{n}$ is the body's surface outward normal vector, $\hat{\bm{t}}$ is the traction given on the Neumann boundary $\varSigma_{\mathrm{N}}$ and $\hat{\bm{u}}$ is the displacement prescribed on the Dirichlet boundary $\varSigma_{\mathrm{D}}$. 
We define $\bm{u}_{\mathrm{gl}}=\bm{u}_0+\bm{u}_{\mathrm{D}}$, with a suitable \textit{dirichlet lift} $\bm{u}_{\mathrm{D}}\in\mathbb{V}_{\mathrm{D}}=\{\bm{v}\in{[H^1(\varOmega_{\mathrm{gl}})]}^d: \bm{v}=\bm{g}_{\mathrm{D}}\;\mathrm{on}\;\Sigma_{\mathrm{D}}\}$ in case of inhomogeneous Dirichlet boundary conditions.
The weak form for \cref{eq:strong_bvp} reads: find $\bm{u}_{0}\in\mathbb{V}=\{\bm{v}\in{[H^1(\varOmega_{\mathrm{gl}})]}^d: \bm{v}=\bm{0}\;\mathrm{on}\;\Sigma_{\mathrm{D}}\}$ such that
\begin{equation}
	\label{eq:weak_form}
	a(\bm{u}_{\mathrm{gl}}, \bm{v}) = f(\bm{v})\,, \quad \forall \bm{v} \in \mathbb{V}\,,
\end{equation}
where
\begin{align}
	a(\bm{w}, \bm{v}) &= \sum_{m=1}^M \int_{\varOmega_{\mathrm{gl}}^m} 
    \lambda^1_m \tr(\varepsilon(\bm{w})) \tr(\varepsilon(\bm{v})) + 
    2\lambda^2_m \varepsilon(\bm{w}) \vdot\!\vdot\, \varepsilon(\bm{v}) \; \dd V
\end{align}
and
\begin{equation}
	f(\bm{v}) = \int_{\varSigma_{\mathrm{N}}} \hat{\bm{t}} \vdot \bm{v}\; \dd A - a(\bm{u}_{\mathrm{D}}, \bm{v})\,.
\end{equation}
Here, $\varOmega_{\mathrm{gl}}^m$ is used to indicate the parts of the global domain associated with the different phases of the heterogeneous material.
Note that, $\tr(\bullet)$ denotes the trace of a tensor and `$\vdot\,\vdot$' stands for the scalar product of two $2$nd-order tensors ($2$-fold contraction as defined in section 2.1.15 of~\cite{BertramGluege2015}).
The energy inner product and energy norm are defined as
\begin{align}
	{\left(\bm{w}, \bm{v}\right)}_{\mathbb{V}} &= a(\bm{w}, \bm{v})\,,\quad \forall \bm{w}, \bm{v} \in\mathbb{V},\\
	\norm{\bm{v}}_{\mathbb{V}}^2 &= a(\bm{v}, \bm{v})\,,\quad \forall \bm{v} \in\mathbb{V}\,.
\end{align}
\subsection{Full order model}%
\label{sub:full_order_model}

The \textit{direct numerical solution} or \textit{full order model} is defined as the finite element approximation of
\cref{eq:weak_form}, searching for the solution in a high fidelity discrete space $\mathbb{V}_{{\delta}}\subset\mathbb{V}$:
find $\bm{u}_{{\delta}} \in \mathbb{V}_{{\delta}}\subset\mathbb{V}$, such that
\begin{equation}
	a({\bm{u}_{{\delta}}}, \bm{v}) = f(\bm{v})\,,\quad \forall \bm{v} \in \mathbb{V}_{{\delta}}\,.
\end{equation}
The dimension of the discrete space is denoted by $N_{\delta} = \dim(\mathbb{V}_{\delta})$, and we denote by ${\{\phi_i\}}_{i=1}^{N_{\delta}}$ a standard finite element basis of $\mathbb{V}$ such that, the stiffness matrix and right hand side can be written as
\begin{equation}
	{(\bm{A}_{\delta})}_{ij} = a(\phi_j, \phi_i)\,,\quad {(\bm{f}_{\delta})}_i = f(\phi_i)\,.
\end{equation}
\section{Multiscale method}%
\begin{figure}[tb]
	\centering
	\includegraphics[width=0.8\linewidth]{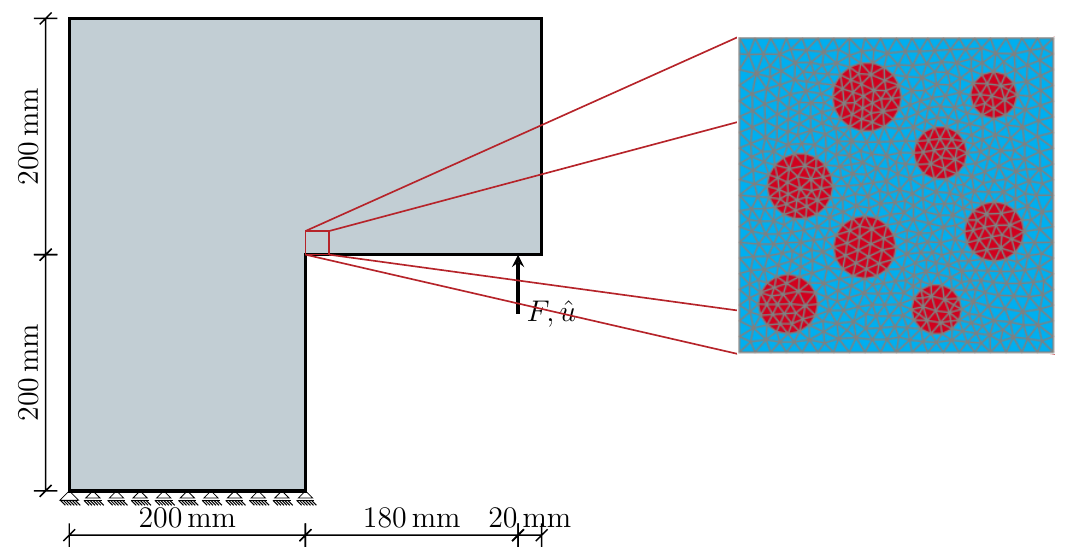}
	\caption{L-shaped panel test specimen and exemplary fine scale structure.}%
	\label{fig:lpanel_specimen_with_rce}
\end{figure}

\label{sec:multiscale_method}
Every material is intrinsically multiscale. 
In the framework of continuum mechanics, often the assumption of a homogeneous material to solve macroscale problems is sufficient.
However, it is only a homogenized approximation of the underlying finer scales and thus not suitable in loading conditions where the real physical phenomena on the fine scale greatly influence the macroscopic behaviour.
Consider for example the propagation of a crack in the test specimen shown in \cref{fig:lpanel_specimen_with_rce}, which is only tractable by resolving the fine scale in the numerical model.
The macroscopic approximation then needs to be improved by fine scale functions taking into account fluctuations in the displacement field due to the heterogeneous fine scale structure.
Note that the discretization of the fine scale structure might vary over the whole mesoscale structure. 

\begin{figure}[tb]
	\centering
	\includegraphics[width=.8\textwidth]{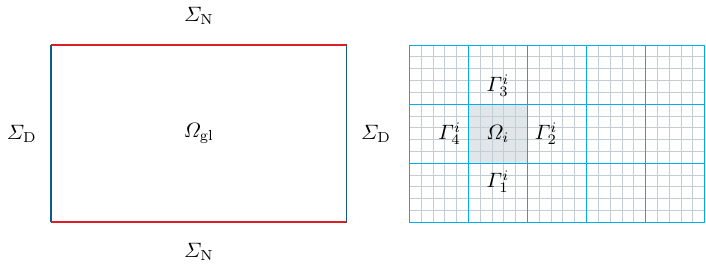}
	\caption{Exemplary computational domain $\varOmega_{\mathrm{gl}}$ on the left and coarse scale (blue lines) and fine scale (gray lines) grid partition on the right. A subdomain $\varOmega_i$ and its edges $\varGamma_e^i$ are also shown.}%
	\label{fig:domain_partitions}
\end{figure}

Following this line of thought, we introduce an additive split of the displacement solution as in the VMM~\cite{Hughes1995,HughesFeijooEtAl1998} $\bm{u}_{0} = \bm{u}_{\mathrm{c}} + \bm{u}_{\mathrm{f}}$ and (in analogy to \cref{sec:problem_setting_and_modeling}) the Hilbert space
\begin{equation}
	\label{eq:split}
	\mathbb{V} = \mathbb{V}_{\mathrm{c}} \oplus \mathbb{V}_{\mathrm{f}}\,,
\end{equation}
is introduced, such that
\begin{align}
    \label{eq:coarse-form}
    a({\bm{u}}_{\mathrm{c}}, \bm{v}_{\mathrm{c}}) + a(\bm{u}_{\mathrm{f}}, \bm{v}_{\mathrm{c}}) &= f(\bm{v}_{\mathrm{c}}) \quad\forall \bm{v}_{\mathrm{c}} \in\mathbb{V}_{\mathrm{c}}\,,\\
    \label{eq:fine-form}
    a({\bm{u}}_{\mathrm{c}}, \bm{v}_{\mathrm{f}}) + a(\bm{u}_{\mathrm{f}}, \bm{v}_{\mathrm{f}}) &= f(\bm{v}_{\mathrm{f}}) \quad\forall \bm{v}_{\mathrm{f}} \in\mathbb{V}_{\mathrm{f}}\,.
\end{align}
Here ${(\bullet)}_{\mathrm{c}}$ denotes the coarse scale part and ${(\bullet)}_{\mathrm{f}}$ the fine scale part. 
The corresponding discrete spaces are associated with coarse scale and fine scale partitions of the domain $\varOmega_{\mathrm{gl}}$ as depicted in \cref{fig:domain_partitions}.
It is assumed that the global domain can be decomposed into $N_{\mathrm{c}}$ non-overlapping
subdomains denoted by $\varOmega_i\,,\; i=1, 2, \ldots, N_{\mathrm{c}}$.
Within each subdomain~$\varOmega_i$, the computational domain for each of the material phases is denoted by $\varOmega_i^m$ with $m=1, \ldots, M$ in analogy to the definitions in \cref{sec:problem_setting_and_modeling}.
Furthermore, the subdomain boundaries are denoted by $\varGamma^i_e$, with $e=1, 2, 3, 4$ referring to the respective bottom, right, top and left edge of the subdomain.
Analogously, the global space $\mathbb{V}$ is decomposed into subdomain spaces $\mathbb{V}_i$ with dimension $n^i_{{\delta}} = \dim(\mathbb{V}_i)$,
\begin{equation}
	\mathbb{V} = \bigoplus_{i=1}^{N_{\mathrm{c}}} \mathbb{V}_i\,.
\end{equation}
\Cref{sub:construction_of_local_approximation_spaces} deals with the construction of reduced basis functions which yield a good approximation of $\bm{u}_{\mathrm{gl}}$, such that the PDE \cref{eq:strong_bvp} is locally fullfilled on $\varOmega_i$.
The definition of the global approximation is described in \cref{sub:reduced_order_model}.

\subsection{Construction of local approximation spaces}%
\label{sub:construction_of_local_approximation_spaces}
In this section, the construction of the reduced basis functions for the coarse scale, ${\{\bm{\varphi}_i\}}_{i=1}^{n_{\mathrm{c}}}$, and for the fine scale, ${\{\bm{\psi}_i\}}_{i=1}^{n_{\mathrm{f}}}$, is addressed.
The proposed \textit{offline phase} consists of three stages:
\begin{enumerate}
	\item direct calculation of coarse scale basis ${\{\bm{\varphi}_i\}}_{i=1}^{n_{\mathrm{c}}}$,
	\item calculation of fine scale edge basis functions ${\{\bm{\chi}^e_i\}}_{i=1}^{n_{\mathrm{mpe}}}$, where $n_{\mathrm{mpe}}$ denotes the number of modes per (subdomain) edge; via two approaches: 
		\begin{enumerate}
			\item by solving an oversampling problem (leading to an \textit{empirical} basis),
			\item by constructing \textit{hierarchical} finite element shape functions ${h}_{p+1}$ (see e.\,g.\, chapter 8 in~\cite{ZienkiewiczTaylor2000}) of degree $p+1$, where $p$ is the polynomial degree of the Legendre polynomial $P_p(\xi)$,
				\begin{equation}
					\label{eq:hierarchical_shapes}
					{h}_{p+1} = \int P_p(\xi)\,\dd\,\xi\,,\quad\mathrm{with}\; P_p(\xi)=\frac{1}{(p-1)!}\frac{1}{2^{p-1}}\frac{\dd^p}{\dd\,\xi^p}[{(\xi^2 -1)}^p]\,,
				\end{equation}
		\end{enumerate}
	\item calculation of fine-scale subdomain basis functions ${\{\bm{\psi}_i\}}_{i=1}^{n_{\mathrm{f}}}$ from ${\{\bm{\chi}^e_i\}}_{i=1}^{n_{\mathrm{mpe}}}$.
\end{enumerate}
Note that, in the remainder of this article, we refer to the set of coarse scale basis functions and fine-scale subdomain basis functions constructed from hierarchical finite element shape functions as the \textit{hierarchical basis}.
The set of coarse scale basis functions and fine scale subdomain basis functions constructed from empirical fine scale edge basis functions is referred to as \textit{empirical basis}.
In the latter case, we study two different \added[id=pd]{sampling strategies to generate} training sets,
\replaced[id=pd]{drawing samples either from a \textit{correlated} or \textit{uncorrelated} sampling distribution}{namely the \textit{multivariate normal} training set and the \textit{normal} training set}, \added[id=pd]{both of} which are defined in~\cref{ssub:training_set}.

\subsubsection{Direct calculation of coarse scale basis functions}%
\label{ssub:coarse_scale_basis}
The coarse scale basis is required to form a partition of unity on the
subdomain boundary $\partial\varOmega_i$, to enable the assembly procedure
described in \cref{sub:reduced_order_model}.
Therefore, we choose piecewise bilinear functions to approximate the coarse scale part,
such that the fine scale part vanishes at the vertices of the coarse scale grid,
similar to the hierarchical shape functions.
By extending standard finite element shape functions on the boundary
of the subdomain into the interior of the respective subdomain,
the effect of the differential operator in the interior is also incorporated
in the coarse scale basis functions.
The coarse scale basis functions are defined as the solution of
\begin{equation}
	\label{eq:coarse_basis_form}
    \added{a_i}(\bm{\varphi}_j, \bm{v}) = 0\,,\quad \bm{\varphi}_j = \bm{\Phi}_j\;\mathrm{on}\;\partial\varOmega_i\,,\quad\forall\bm{v}\in\mathbb{V}_i\,,
\end{equation}
where
\begin{align}
	\label{eq:a_rce}
    \added{a_i}(\bm{w}, \bm{v}) &= \sum_{m=1}^M \int_{\varOmega^m_i} 
    \lambda^1_m \tr(\varepsilon(\bm{w})) \tr(\varepsilon(\bm{v})) + 
    2\lambda^2_m \varepsilon(\bm{w}) \vdot\!\vdot\, \varepsilon(\bm{v}) \; \dd V
\end{align}
and $\bm{\Phi}_j$ being the standard Lagrange basis functions, which are constructed from Lagrange ploynomials (see~\cite{ZienkiewiczTaylor2000})
\begin{equation}
	{l}^{q}_{k}(\xi)=\frac{%
		(\xi-{\xi}_{0})(\xi-{\xi}_{1})\cdots(\xi-{\xi}_{k-1})(\xi-{\xi}_{k+1})\cdots(\xi-{\xi}_{q})
	}{%
	({\xi}_{k}-{\xi}_{0})({\xi}_{k}-{\xi}_{1})\cdots({\xi}_{k}-{\xi}_{k-1})({\xi}_{k}-{\xi}_{k+1})\cdots({\xi}_{k}-{\xi}_{q})
}\,,
\end{equation}
giving unity at ${\xi}_{k}$ and passing through $q$ points.
In two dimensions, the node $j$ of the coarse grid element may be labeled by its column and row number, $I$, $J$,
\begin{equation}
	{\Phi}_{j}={l}^{q}_{I}(\xi){l}^{r}_{J}(\eta)\,.
\end{equation}
The integer $q$ and $r$ stand for the number of subdivisions in each direction and $\xi$ and $\eta$ for the reference coordinates.
For quadrilateral coarse grid cells in the two dimensional case and linear interpolation
in the coarse scale, this yields a local coarse scale basis of size $n_{\mathrm{c}}=8$,
which is used in the remainder of the article.

\subsubsection{Calculation of fine scale edge basis functions}%
\begin{figure}[tb]
\begin{center}
	\includegraphics[keepaspectratio]{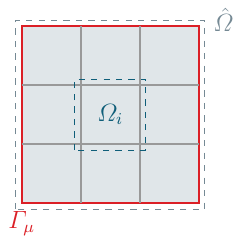}
\end{center}
\caption{Oversampling domain $\hat{\varOmega}$, target subdomain $\varOmega_i$, and the boundary $\varGamma_{\mu}$ for a target subdomain in the interior of the structure are shown.}%
\label{fig:oversampling}
\end{figure}

\label{ssub:fine_scale_edge_basis}
As mentioned in the beginning of \cref{sub:construction_of_local_approximation_spaces}, two different approaches are considered. 
The fine scale edge basis functions may be defined as \textit{hierarchical} shape functions a priori. 
In this case, no precomputation is required and one could directly compute the extension of the edge basis into the interior of the subdomain, as described in 
\cref{ssub:fine_scale_subdomain_basis}.
The construction of the \textit{empirical} fine scale edge basis poses the main challenge in the proposed framework. 
In order to exploit fine scale solutions of the PDE~\cref{eq:strong_bvp} on any subdomain $\varOmega_i$, we make use of the concept of oversampling~\cite{HW1997}.
First, the oversampling domain $\hat{\varOmega}$ is defined, such that $\varOmega_{i}\subsetneq\hat{\varOmega}\subset\varOmega_{\mathrm{gl}}$.
Furthermore, the distance between the boundary $\partial\varOmega_{i}$ and $\varGamma_{\mu}:=\partial\hat{\varOmega}\setminus\partial\varOmega_{\mathrm{gl}}$ is greater than zero, i.\,e.\ $\dist(\varGamma_{\mu}, \partial\varOmega_{i})\ge\rho>0$ for some $\rho$.
Depending on the configuration for a particular $\varOmega_i$, $\varGamma_{\mathrm{N}}:=\partial\hat{\varOmega}\cap\varSigma_{\mathrm{N}}$ or $\varGamma_{\mathrm{D}}:=\partial\hat{\varOmega}\cap\varSigma_{\mathrm{D}}$ may be not empty and Neumann or Dirichlet boundary conditions of the global problem need to be considered in the oversampling as well.
In order to sufficiently incorporate Dirichlet and Neumann boundary conditions in the reduced basis functions, several oversampling problems have to be defined.
We also refer to the different oversampling problems as \enquote{configurations} due to the possible change in topology and boundary conditions for each problem.
The challenge in solving \cref{eq:strong_bvp} on $\hat{\varOmega}$ lies in the definition of the boundary data on $\varGamma_{\mu}$ which is used to exploit possible solutions of the PDE on $\varOmega_{i}$.
Taking, for example, parametric boundary conditions on $\varGamma_{\mu}$, for a specific numerical discretization, the maximum size of the parameter space $\mathbb{P}$ is the number of degrees of freedom on $\varGamma_{\mu}$.
Consider for example $\hat{\varOmega}$ as a $3\times 3$ block of mesoscale subdomains, as shown in~\cref{fig:oversampling}.
With the subdomain type I, discretized with 11 vertices per edge, as shown in~\cref{fig:rce_types_a}, this leads to 120 vertices on $\varGamma_{\mu}$.
For linear triangular elements, this would lead to a parameter space $\mathbb{P} = \mathbb{R}^{240}$, where a dense uniform sampling as usually done in standard greedy approaches (see~\cite{VeroyEtAl2003}) is infeasible.
For this reason, Buhr and Smetana~\cite{BS2018} suggest to solve~\cref{eq:strong_bvp} on $\hat\varOmega$ with random boundary conditions on $\varGamma_{\mu}$ (i.\,e.\,, the associated transfer operator $\bm{T}$ is approximated by random sampling).
Note, that the transfer operator $\bm{T}$ maps functions on the boundary $\varGamma_{\mu}$ to the solution of the PDE on the target subdomain $\varOmega_i$.
As an example, \cref{fig:oversampling} shows the oversampling domain $\hat{\varOmega}$ for a subdomain of interest $\varOmega_i$ that is entirely inside the structure.

In this work, we adopt \enquote{Algorithm 1: Adaptive Randomized Range Approximation}~\cite{BS2018} and modify it to include the solution of a global reduced problem in the training data.
In \cref{sec:numerical_experiments}, we then compare the randomized approach using an \replaced[id=pd]{\textit{uncorrelated} sampling strategy}{\textit{normal} distribution with zero mean} with the proposed \added[id=pd]{\textit{correlated}} sampling \added[id=pd]{strategy}\deleted[id=pd]{from a \textit{multivariate normal} distribution with the solution of the global reduced problem as mean}.
More details on the different sampling strategies are discussed in~\cref{ssub:training_set}.
For ease of notation, we assume that \added[id=pd]{there exists a} suitable training set $S_{\mathrm{train}}$ containing samples \replaced[id=pd]{generated using}{drawn from} either \deleted[id=pd]{one }of the aforementioned
\replaced[id=pd]{sampling strategies}{distributions}\deleted{ exists}.

The \textit{oversampling problem}
\begin{equation}
	\label{eq:oversampling_problem}
    \begin{aligned}
        - \nabla\vdot\bm{\sigma}(\nabla\bm{u}) &= 0 \quad &&\mathrm{in}\;\hat{\varOmega}_{}\,,\\
        \bm{\sigma}(\nabla\bm{u})\vdot\bm{n} &= \bm{0} \quad &&\mathrm{on}\; \varGamma_{\mathrm{N}}\,,\\
        \bm{u} &= \bm{0} \quad &&\mathrm{on}\;\varGamma_{\mathrm{D}}\,,\\
	    \bm{u} &= \bm{g} \quad &&\mathrm{on}\;\varGamma_{\mu}\,,\forall \bm{g}\in S_{\mathrm{train}}\,,
    \end{aligned}
\end{equation}
is then solved for $\bm{u}$ for each element $\bm{g}\in S_{\mathrm{train}}$ in the training set prescribed on the boundary $\varGamma_{\mu}$.
In case of inhomogeneous Dirichlet data $\bm{g}_{\mathrm{D}}$ on $\varSigma_{\mathrm{D}}$, the fine scale part $\bm{g}_{\mathrm{f}} = \bm{g}_{\mathrm{D}} - \bm{g}_{\mathrm{c}}$ is set as the basis for the respective edges.
The coarse scale part
$\bm{g}_{\mathrm{c}}= \sum_{i=1}^{n_{\mathrm{c}}}\bm{g}_j \bm\varphi_i$
is given as a linear combination of the function value of $\bm{g}$ at the
$j$-th vertex of the coarse grid cell and the coarse scale basis functions
defined in~\cref{ssub:coarse_scale_basis}.
In case of non-zero Neumann data $\hat{\bm{t}}\ne \bm{0}$,
the fine scale edge basis is extended by
additional edge functions obtained by solving the problem
\begin{equation}
	\label{eq:oversampling_problem_neumann}
	\begin{aligned}
		- \nabla\vdot\bm{\sigma}(\nabla\bm{u}) &= 0 \quad &&\mathrm{in}\;\hat{\varOmega}_{}\,,\\
		\bm{\sigma}(\nabla\bm{u})\vdot\bm{n} &= \hat{\bm{t}} \quad &&\mathrm{on}\; \varGamma_{\mathrm{N}}\,,\\
		\bm{u} &= \bm{0} \quad &&\mathrm{on}\;\varGamma_{\mu}\,.
	\end{aligned}
\end{equation}

The computational procedure to construct the fine scale edge basis functions
which consists of repeatedly solving~\cref{eq:oversampling_problem} for different
training samples is summarized in~\cref{algo:rrf}.

Following Buhr and Smetana~\cite{BS2018}, we let $\mathcal{P}$ be the probability, $P_{\mathrm{span}(\bm{B})}$ the orthogonal projection onto $\mathrm{span}(\bm{B})$, $N_{T}$ the rank of the transfer operator $\bm{T}$ and $\lambda^{M_S}_{\mathrm{min}}$ the smallest eigenvalue of the matrix of the inner product in the source space of the transfer operator.
First, a set of edge basis functions $\bm{B}_e$ and $n_t$ testvectors $\bm{M}_e$
for each edge $e$ of the target subdomain $\varOmega_i$ are initialized.
The test data consists in solutions of \cref{eq:oversampling_problem}
with uncorrelated random samples $\bm{g}$ \added[id=pd]{(drawn from a multivariate normal distribution with zero mean and covariance matix $\bm\varSigma=\mathds{1}$)} prescribed as boundary data
on $\varGamma_{\mu}$.
The restriction of the solution to the target subdomain $\varOmega_i$ is equivalent
to the image of the transfer operator $\bm{T}$ and hence the restriction
of the solution to one of the edges of the target subdomain is
denoted as $\bm{T}\bm{g}\vert_{\varGamma_e^i}$.
For each of the testvectors in $\bm{M}_e$ the coarse scale part is subtracted.
Next, the error estimator factor $c_{\mathrm{est}}$ is determined
based on the chosen inner products.
The body of the while loop consists in drawing a new sample
\replaced[id=pd]{using the chosen sampling strategy}{from the chosen distribution} (either \replaced[id=pd]{\textit{uncorrelated}}{\textit{normal}} or \replaced[id=pd]{\textit{correlated}}{\textit{multivariate
normal}}), computing the solution to \cref{eq:oversampling_problem}
and restricting it to the edges of the target subdomain.
Subsequently, the coarse scale part is subtracted and each edge basis function
set $\bm{B}_e$ is extended \added[id=pd]{by adding the fine scale part of the solution
and orthonormalizing the edge basis using the modified Gram-Schmidt algorithm.}
In the last step, the set of testvectors for each edge is orthogonalized
with respect to the edge basis set.
This way, with increasing number of iterations the norm of each of the
testvectors in the test sets $\bm{M}_e$ is decreased until the
criterion to terminate the while loop is met.
If the criterion for one of the edges is met before the others,
the algorithm is continued without adding basis functions for
that particular edge.
\deleted{It is noted that the POD algorithm~\cite{pymor} used in this paper computes the POD modes via the \textit{method of snapshots}~\cite{Sirovich1987} which is summarized briefly in the paragraph below.}
The fine scale edge basis is denoted as ${\{\bm{\chi}^e_j\}}_{j=1}^{n^e_{\mathrm{mpe}}}$, with $e=1, 2, 3, 4$ referring to one of the edges of the subdomain $\varGamma_e^i$ and $n^e_{\mathrm{mpe}}$ denoting the number of modes per edge for a particular edge $e$.

\begin{algorithm}[ht]
\caption{Modified Adaptive Randomized Range Approximation}%
\label{algo:rrf}
\begin{algorithmic}[1]
    \Function{AdaptiveRandRangeApprox}{$\bm{T}$, $\texttt{tol}$, $n_t$, $\varepsilon_{\mathrm{algofail}}$}

    \textbf{Input:} Operator $\bm{T}$, target tolerance $\texttt{tol}$, number of testvectors $n_t$, maximum failure probability $\varepsilon_{\mathrm{algofail}}$, training set $S_{\mathrm{train}}$

    \textbf{Output:} Fine scale edge basis $\bm{B}_e$ for each of the edges $\varGamma_{e}^i$ of the target subdomain $\varOmega_{i}$ where each edge basis fulfills $R^n=\mathrm{span}(\bm{B}_e)$ with property $\mathcal{P} \left(\norm{\bm{T} - P_{R^n}\bm{T}}\le\texttt{tol}\right) > (1 - \varepsilon_{\mathrm{algofail}})$
    \State{$\bm{B}_e$ $\gets\;\emptyset$}\Comment{initialize basis for each edge}
    \State{$\bm{M}_e\gets\{ \bm{T}\bm{g}^1\vert_{\varGamma_e^i}, \ldots, \bm{T}\bm{g}^{n_t}\vert_{\varGamma_e^i}\}$}\Comment{initialize test vectors for each edge}
    \State{$\bm{M}_e\gets \bm{M}_e - \bm{M}_e^{\mathrm{c}}$}\Comment{subtract coarse scale part}
    \State{$\varepsilon_{\mathrm{testfail}}\gets\varepsilon_{\mathrm{algofail}}/N_{\bm{T}}$}
\State{$c_{\mathrm{est}}\gets{\left[\sqrt{2\lambda^{M_S}_{\min}}\mathrm{erf}^{-1}\left(\sqrt[n_t]{\varepsilon_{\mathrm{testfail}}}\right)\right]}^{-1}$}\Comment{determine error estimator factor}
\While{$\left(\max_{t\in \bm{M}_e} \norm{t}_R\right) \vdot c_{\mathrm{est}} > \texttt{tol}$}\Comment{compare maxnorm to target tol for each test set}
	\State{$\bm{g}\gets$ draw sample from training set $S_{\mathrm{train}}$}
    \State{$\bm{u}\gets \bm{T}\bm{g}\vert_{\varGamma_e^i}$}\Comment{restriction of the solution to each edge}
	\State{$\bm{u}_{\mathrm{f}} \gets \bm{u} - \bm{u}_{\mathrm{c}}$}
    \State{$\bm{B}_e\gets \bm{B}_e\cup(\bm{u}_{\mathrm{f}})$}
    \State{$\bm{B}_e\gets \added{\mathrm{orthonormalize}}\,\left(\bm{B}_e\right)$}
    \State{$\bm{M}_e\gets\{t-P_{\mathrm{span}(\bm{B}_e)}t \mid t\in \bm{M}_e\}$}\Comment{orthogonalize test vectors to $\mathrm{span}(\bm{B}_e)$}
\EndWhile{}
\State{\textbf{return} {$\bm{B}_e$}}
\EndFunction%
\end{algorithmic} 
\end{algorithm}


\subsubsection{Calculation of fine scale subdomain basis functions}%
\label{ssub:fine_scale_subdomain_basis}
The final step in constructing the reduced basis for the fine scale part of the displacement solution consists in the extension of the \replaced[id=pd]{edge basis functions}{POD edge modes} into the respective subdomain $\varOmega_i$.
It is important to note that, by setting the same function on a single edge
for adjacent subdomains, continuity of the global approximation is ensured.
Furthermore, to enable the standard assembly procedure as in the finite element method,
in the local extension problem, it is necessary to enforce zero boundary
conditions on edges
$\partial\varOmega_{i}\setminus\varGamma_{e}^{i}$ where the \replaced{basis function}{POD
mode} is not prescribed.
For each edge $\varGamma_{e}^{i}$, with $e=1, 2, 3, 4$, and each \deleted{POD} mode
in the set of edge basis functions
${\{\bm{\chi}_j^e\}}_{j=1}^{n^e_{\mathrm{mpe}}}$,
find the subdomain basis function $\bm{\psi}_j^e$ such that
\begin{align}
	\label{eq:fine_basis_form}
	\begin{split}
        \added{a_i}(\bm{\psi}_j^e, \bm{v}) &= 0\,,\quad \quad\forall\bm{v}\in\mathbb{V}_i\,,\\
	\mathrm{with}\quad\bm{\psi}_j^e &= \bm{\chi}_j^e\;\mathrm{on}\;\varGamma_e^i\,\quad\mathrm{and}\quad \bm{\psi}_j^e = \bm{0}\;\mathrm{on}\;\partial\varOmega_i\setminus\varGamma_e^i\,,
	\end{split}
\end{align}
where the bilinear form $\added{a_i}(\cdot, \cdot)$ is given by \cref{eq:a_rce}.
All $n_{\mathrm{f}} = \sum_{e=1}^{4} n^e_{\mathrm{mpe}}$ solutions of \cref{eq:fine_basis_form} are then gathered in one set of fine scale basis functions 
\begin{equation}
	{\{\bm{\psi}_j\}}_{j=1}^{n_{\mathrm{f}}} = {\{\bm{\psi}_k^1\}}_{k=1}^{n^1_{\mathrm{mpe}}}\cup{\{\bm{\psi}_k^2\}}_{k=1}^{n^2_{\mathrm{mpe}}}\cup{\{\bm{\psi}_k^3\}}_{k=1}^{n^3_{\mathrm{mpe}}}\cup{\{\bm{\psi}_k^4\}}_{k=1}^{n^4_{\mathrm{mpe}}}\,.
\end{equation}

\subsubsection{Sampling distributions}%
\label{ssub:training_set}
In this section, the two different choices (sampling \replaced[id=pd]{strategies}{distributions} considered) for the definition of the training set
$S_{\mathrm{train}}$ mentioned in~\cref{ssub:fine_scale_edge_basis} are discussed.

In the first variant, \replaced[id=pd]{\textit{uncorrelated} samples are}{a \textit{normal} distribution is} used to define
the random boundary conditions in the oversampling problem.
Following Buhr and Smetana~\cite{BS2018}, we use the term \enquote{random normal vector}
to \enquote{denote a vector whose entries are independent and identically
    distributed random variables with normal distribution}.
Each entry $g_i$ of the vector $\bm{g}\in S_{\mathrm{train}}$ is sampled from a
normal distribution with zero mean and variance of one, which is denoted as
\begin{equation}
	g_i \sim\mathcal{N}(0, 1)\,.
\end{equation}
\added[id=pd]{Note that throughout the manuscript, we use the term \textit{uncorrelated} to refer to
the random samples which are drawn from a normal distribution with
zero mean and variance of one.}

In the second variant, \added[id=pd]{\textit{correlated}} samples are drawn from a \textit{multivariate normal} distribution
with the solution of the global reduced problem $\bm{u}_{\mathrm{c}}$ as mean.
The solution $\bm{u}_{\mathrm{c}}$ is obtained by solving the reduced order model
(see~\cref{sub:reduced_order_model})
using only the coarse scale basis.
\begin{equation}
	\bm{g}\sim\mathcal{N}(\bm{u}_{\mathrm{c}}, \bm{\varSigma})\,.
\end{equation}
Here, $\bm{\varSigma}$ denotes the covariance matrix of the distribution.
The covariance matrix 
\begin{equation}
	\bm{\varSigma} = \bm{M}\bm{P}\bm{M}
\end{equation}
is computed based on the matrix $\bm{P}$ with elements $p_{ij}= \exp(-d_{ij}/L_{\mathrm{corr}})$ which
defines the correlation of two entries $g_i$ and $g_j$ of $\bm{g}$ based on the euclidean distance $d_{ij}$ between the points associated with the entries in $\bm{g}$ and the matrix $\bm{M} = \mathrm{diag}(\bm{u}_{\mathrm{c}})$.
$L_{\mathrm{corr}}$ is the correlation length and controls the strength of correlation
between two points.
\added[id=pd]{Note that throughout the manuscript, we use the term \textit{correlated} to refer to
the random samples, which are drawn from a multivariate normal distribution with
$\bm{u}_{\mathrm{c}}$ as mean and covariance matrix $\bm\varSigma$ as described above.}

\added[id=pd]{In the limit $L_{\mathrm{corr}}\to 0$, the correlation matrix $\bm{P}$
becomes the identity $\mathds{1}$, and hence the same result as training the basis with the fully uncorrelated samples is to be expected.
Therefore, for the examples presented in~\cref{sec:numerical_experiments}, the correlation length is first set to the maximum Euclidean distance between two points of the oversampling domain.
This effectively includes the correlated, smoother macroscopic states in the training.
The number of eigenvalues of $\bm{\varSigma}$ whose values are greater than
$5\%$ of the largest eigenvalue, is taken as the number
of samples to be drawn using this correlation length.
The tolerance on the eigenvalues, which results in limiting the number of samples per correlation length, is used to prevent drawing fully dependent samples. 
Subsequently, the correlation length is halved
and the number of samples to be drawn using the updated correlation length is
determined based on the relative tolerance on the number of eigenvalues
of $\bm{\varSigma}$ minus the number
of already drawn samples.
This way, the strength of the correlation decreases with increasing number of samples
drawn in the range finder algorithm~(\cref{algo:rrf}).
By decreasing the correlation length, we aim at drawing as many less correlated samples as necessary to achieve a sufficiently good approximation of the range of the transfer operator, as would have been the case when using fully uncorrelated samples
from the start.}

\subsection{Reduced order model}%
\label{sub:reduced_order_model}
The proposed approach features a local basis ${\{\bm{\xi}_k\}}_{k=1}^n = {\{\bm{\varphi}_i\}}_{i=1}^{n_{\mathrm{c}}} \cup{\{\bm{\psi}_j\}}_{j=1}^{n_{\mathrm{f}}}$, with $n$ being the maximum number of basis functions, which constitutes a partition of unity for all vertices of the coarse grid and is continuous on subdomain boundaries $\varGamma_e^i$.
Here $\bm{\varphi}_i$ and $\bm{\psi}_j$ are the $n_{\mathrm{c}}$ coarse scale and $n_{\mathrm{f}}$ fine scale functions, respectively.
The local reduced basis can be expressed in the standard finite element basis $\phi_j$ associated with the fine grid
\begin{equation}
	\bm{\xi}_k = \sum_{j=1}^{n_{\delta}} \bm{B}_{jk} \phi_j\,,
\end{equation}
where the $k$-th column of the matrix $\bm{B}\in\mathbb{R}^{n_{\delta}\times n}$ holds the coefficients of the $k$-th basis function.
The local contribution of a subdomain is then given by
\begin{equation}
	\bm{A}_{n} = \bm{B}^{\mathrm{T}} \bm{A}^{\mathrm{loc}}_{\delta} \bm{B}\,,\quad \bm{f}_{n} = \bm{B}^{\mathrm{T}}\bm{f}^{\mathrm{loc}}_{\delta}\,,
\end{equation}
where $\bm{A}^{\mathrm{loc}}_{\delta}\in\mathbb{R}^{n_{\delta}\times n_{\delta}}$ and $\bm{f}^{\mathrm{loc}}_{\delta}\in\mathbb{R}^{n_{\delta}}$ denote the stiffness matrix and external force vector, respectively, of a subdomain.
Note that unless body forces are present, $\bm{f}^{\mathrm{loc}}_{\delta}$ is zero in case $\varGamma^i_e\cap\varSigma_{\mathrm{N}}=\emptyset$ or $\hat{\bm{t}}=\bm{0}$ which is the case for most subdomains. 
Due to the above-mentioned properties of the reduced basis functions, the local contributions $\bm{A}_{n}\in\mathbb{R}^{n\times n}$ and $\bm{f}_{n}\in\mathbb{R}^{n}$ can be sorted into global vectors following the usual assembly procedure of standard finite elements, where each node, edge and face (3D) is associated with a fixed number of degrees of freedom (DoFs).
The global system of the \textit{reduced order model} of size $N\ll N_{\delta}$ is then written as
\begin{equation}
	\label{eq:rom}
	\bm{A}_{N} \bm{u}_{N} = \bm{f}_N\,,
\end{equation}
where $\bm{A}_{N}\in\mathbb{R}^{N\times N}$, $\bm{f}_{N}\in\mathbb{R}^{N}$ and $N$ is the number of unknown DoFs in the reduced order model.

\section{Numerical Experiments}%
\label{sec:numerical_experiments}
In this section, the performance of the empirical bases, using either the \replaced[id=pd]{uncorrelated samples}{\textit{normal} distribution} or the \replaced[id=pd]{correlated samples}{\textit{multivariate normal} distribution}~(see \cref{ssub:training_set}), is studied and compared to the hierarchical basis as a na\"{\i}ve choice for the approximation of the fine scale part.
\added[id=pd]{We also note that a comparison of Legendre basis functions and an empirical basis was carried out by Eftang and Patera in the context of port reduction for static condensation procedures~\cite{eftang2013port}.}
In \cref{sub:block_example}, a block example is implemented to illustrate the basic features of the proposed methodology.
Next, a beam under the state of pure bending is analyzed for varying ratios of the elastic moduli as a measure for the heterogeneity, and the empirical basis' performance is shown to be superior to that of the hierarchical basis for ratios greater than one.
The applicability of the method to more complex problems (containing a stress singularity in this case) is demonstrated by the example of an L-shaped panel in \cref{sub:l_panel_example}.
Finally, details on the basis construction and the computational time of the FOM as well as offline and online phase of the ROM are given in \cref{sub:computational_time}.

\begin{figure}[htb]
	\centering
	\subfloat[\normalsize Mesoscale subdomain type I\label{fig:rce_types_a}]{\includegraphics[width=0.4\linewidth]{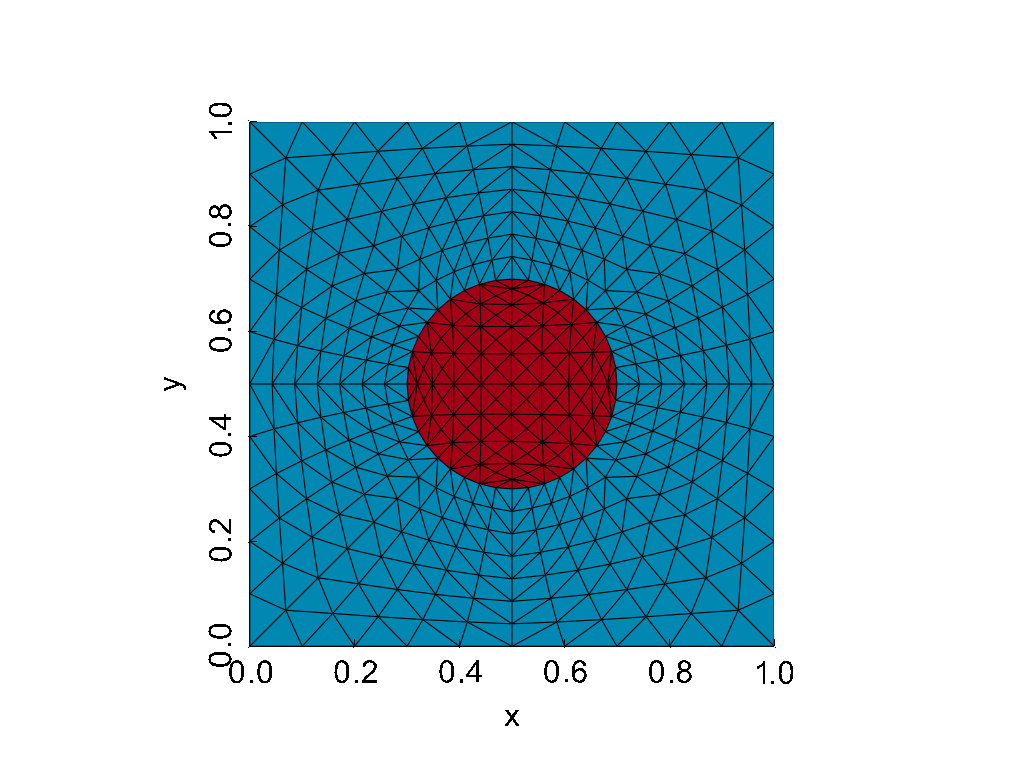}}
	\subfloat[\normalsize Mesoscale subdomain type II\label{fig:rce_types_b}]{\includegraphics[width=0.4\linewidth]{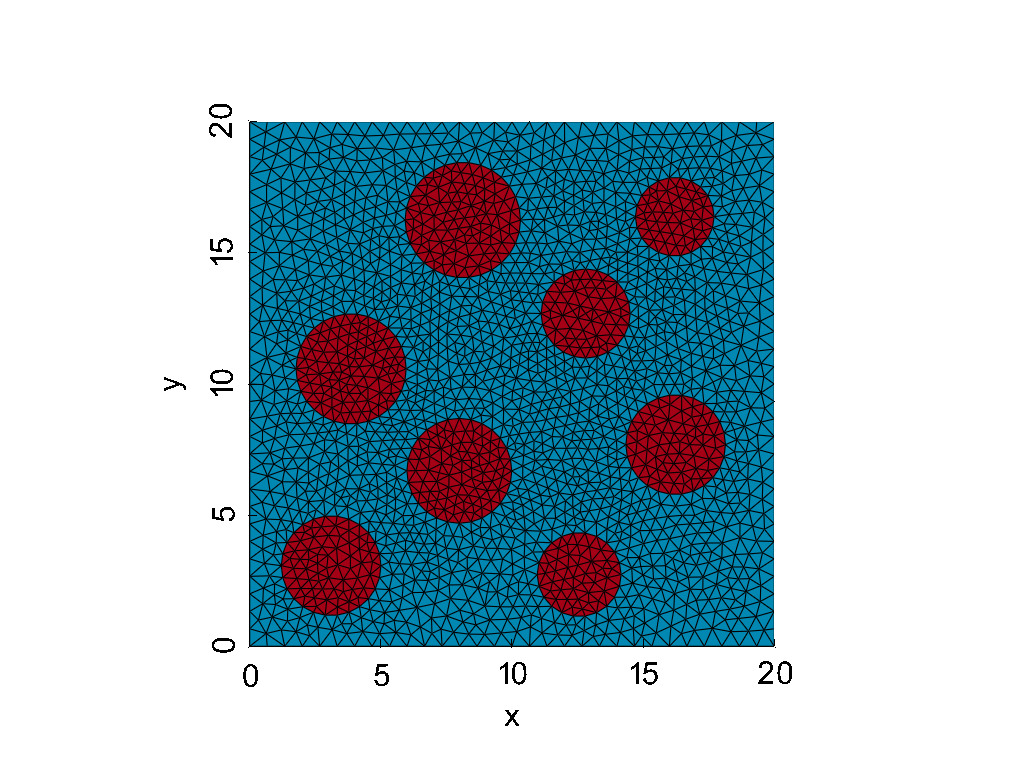}}
	\caption{Mesoscale structures used in the examples.}%
	\label{fig:rce_types}
\end{figure}

\begin{figure}[htb]
	\centering
	\includegraphics[width=0.6\linewidth]{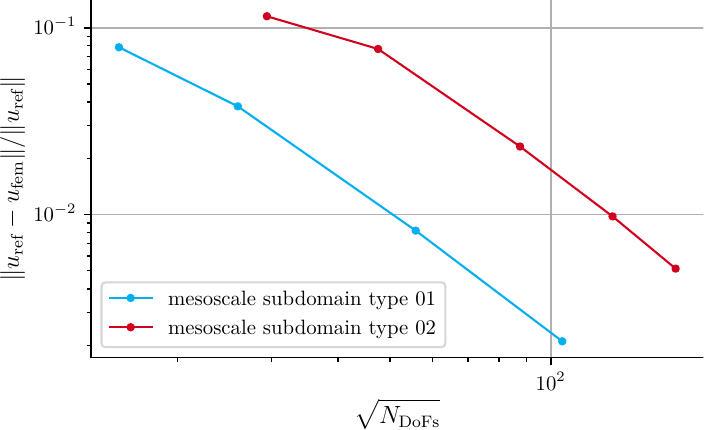}
	\caption{Relative error in the energy norm against square root of number of degrees of freedom in the mesh convergence analysis.}%
	\label{fig:convergence}
\end{figure}

The material parameters are given in \cref{tab:concrete_parameters}.
Triangular elements with quadratic shape functions are implemented for the fine grid discretization of the mesoscale subdomain types used in the examples, as shown in~\cref{fig:rce_types}.
\Cref{fig:convergence} shows the results of the mesh convergence analysis
carried out for the mesoscale subdomains studied in the examples.
For different levels of refinement, \cref{eq:weak_form} is solved on the mesoscale subdomain with boundary data given by \cref{eq:block_boundary_data}, and the error relative to a reference solution computed on the finest mesh is measured in the energy norm.
The mesh is regarded as converged if the relative error in the energy norm is below one percent which leads to the discretizations as shown in~\cref{fig:rce_types}.

In all examples, the global error relative to the \textit{full order model} (\cref{eq:weak_form}) is computed as follows.
The absolute error on subdomain $\varOmega_i$ defined as $\bm{e}_i = {(\bm{u}_{\mathrm{fom}})}_i - {(\bm{u}_{\mathrm{rom}})}_i$ is measured in the $H^1$-norm as
\begin{equation}
	\label{eq:norm_squared}
    \norm{\bm{e}_i}^2_{\mathbb{V}_i} = \int_{\varOmega_i} \bm{e}_i \vdot \bm{e}_i + \nabla\bm{e}_i \vdot\!\vdot\,\nabla\bm{e}_i \,\dd V\,.
\end{equation}
The global absolute error is thus given by the square root of the sum of the squared local norm
\begin{equation}
	\label{eq:gl_abs_err}
	\norm{\bm{e}}_{\mathbb{V}} = \sqrt{\sum_{i=1}^{N_{\mathrm{c}}} \norm{\bm{e}_i}^2_{\mathbb{V}_i}}\,.
\end{equation}
Analogously, the global relative error is given by
\begin{equation}
	\label{eq:gl_rel_err}
	\frac{\norm{\bm{e}}_{\mathbb{V}}}{\norm{\bm{u}_{\mathrm{fom}}}_{\mathbb{V}}}
	= \frac{\sqrt{\sum_{i=1}^{N_{\mathrm{c}}} \norm{\bm{e}_i}^2_{\mathbb{V}_i}}
	}{\sqrt{\sum_{i=1}^{N_{\mathrm{c}}} \norm{{(\bm{u}_{\mathrm{fom}})}_i}^2_{\mathbb{V}_i}}
}\,.
\end{equation}
Furthermore, in each of the examples statistics over \replaced[id=pd]{$\NumReal{}$}{$10$}
realizations for each of the sampling approaches are given.
\begin{table}[tb]
    \centering
    \caption{Material parameters (taken from table 4 in~\cite{Unger2011a}).}%
    \label{tab:concrete_parameters}
    \begin{tabular}{lcc}\toprule
    & Mortar matrix & Aggregates\\ \midrule
	    Young's modulus & $E_{\mathrm{m}} = 30\,000\,\mathrm{MPa}$ & $E_{\mathrm{a}}=60\,000\,\mathrm{MPa}$ \\
	    Poisson ratio & $\nu_{\mathrm{m}} = 0.2$ & $\nu_{\mathrm{a}}=0.2$ \\
       \bottomrule
    \end{tabular}
\end{table}


\FloatBarrier%
\subsection{Block example}%
\begin{figure}[tb]
	\centering
	\includegraphics[width=0.4\linewidth]{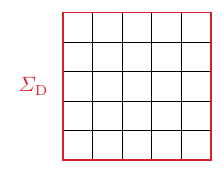}
	\caption{Illustration of the coarse grid discretization of the block example.}%
	\label{fig:block_sketch}
\end{figure}

\label{sub:block_example}
In this example,~\cref{eq:strong_bvp} is solved on a global domain $\varOmega_{\mathrm{gl}} = {(0, 5)}^2$ with Dirichlet data on the boundary $\varSigma_{\mathrm{D}}:= \partial\varOmega_{\mathrm{gl}}$ given in index notation by
\begin{equation}\label{eq:block_boundary_data}
	u_i^{\mathrm{D}} = a_{ij} x_j + b_{ij} x_j^2\,,\quad a_{ij}, b_{ij}\in\mathbb{R}\,.
\end{equation}
The coefficients $a_{ij}$ and $b_{ij}$ are random variables sampled from a uniform
distribution over $[0, 1)$ and scaled such that
$\norm{\bm{u}^{\mathrm{D}}(x_1=5, x_2=5)}=1$.
The mesoscale subdomain of type I (\cref{fig:rce_types_a}) is used for each of the $25$ subdomains, consequently, the coarse grid is chosen as a structured grid with five cells in each spatial direction, as shown in~\cref{fig:block_sketch}.

\begin{figure}[tb]
	\centering
	\includegraphics[width=0.8\linewidth]{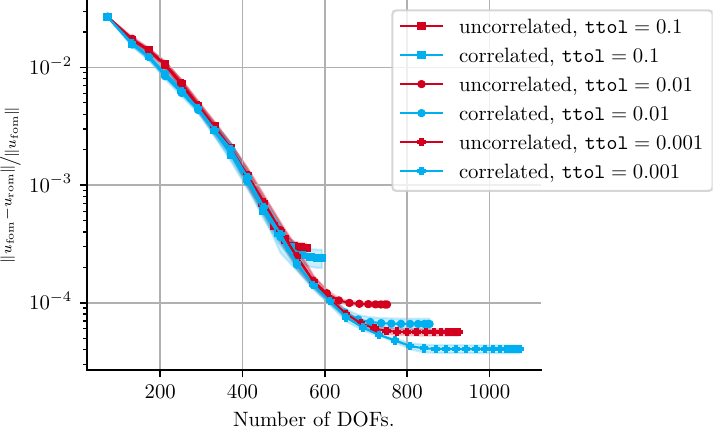}%
    \caption{Block example: \added[id=pd]{global} relative error in the $H^1$-norm
    for different sampling distributions \deleted[id=pd]{(normal and multivariate normal (mvn)) }
    and target tolerances (\texttt{ttol}).
    \added[id=pd]{The values are averaged over the number of $\NumReal{}$ realizations.}
    \added[id=pd]{The shaded areas indicate the standard deviation of the global relative error.}
}%
	\label{fig:block_relerr}
\end{figure}

While the performance of the empirical basis using the \replaced[id=pd]{correlated samples}{\textit{multivariate normal} distribution} is compared to the empirical basis using the \replaced[id=pd]{uncorrelated samples}{\textit{normal} distribution}, the target tolerance as input to the range finder algorithm~\cref{algo:rrf} is varied.
The decay of the relative global error against the number of degrees of freedom in the ROM as shown in~\cref{fig:block_relerr}, is computed as follows.
For a given target tolerance, the fine scale basis functions are computed by
solving $N_{\mathrm{c}}$ (number of subdomains) oversampling problems.
Then the ROM is repeatedly evaluated and compared to
the FOM while the number of fine scale basis
functions per edge is increased (if possible) until the maximum number
of basis functions is reached.
The global relative ROM error is calculated as described
by~\cref{eq:norm_squared,eq:gl_abs_err,eq:gl_rel_err}
in~\cref{sec:numerical_experiments}\added[id=pd]{.}

In this example, the difference between the two sampling approaches is small,
as the curves lie almost on top of each other.
In both cases the error stagnates with increasing number of
DoFs which is due to the fact that the maximum
number of basis functions varies per edge.
The accuracy of the global approximation is thus limited by the worst approximation
of the solution on one of the edges (presumably the edge with the lowest
number of basis
functions generated by~\cref{algo:rrf}).
Moreover, as stated in~\cite{BS2018}, depending on the size of the oversampling domain and
the choice of the inner products, the a \replaced[id=pd]{posteriori}{priori}
error bound can be rather pessimistic and more basis functions than needed
to achieve the target tolerance are generated.

\begin{figure}[htb]
	\centering
    \subfloat[{\normalsize{Uncorrelated sampling}\label{fig:block_max_modes_normal}}]{\includegraphics[width=0.5\linewidth]{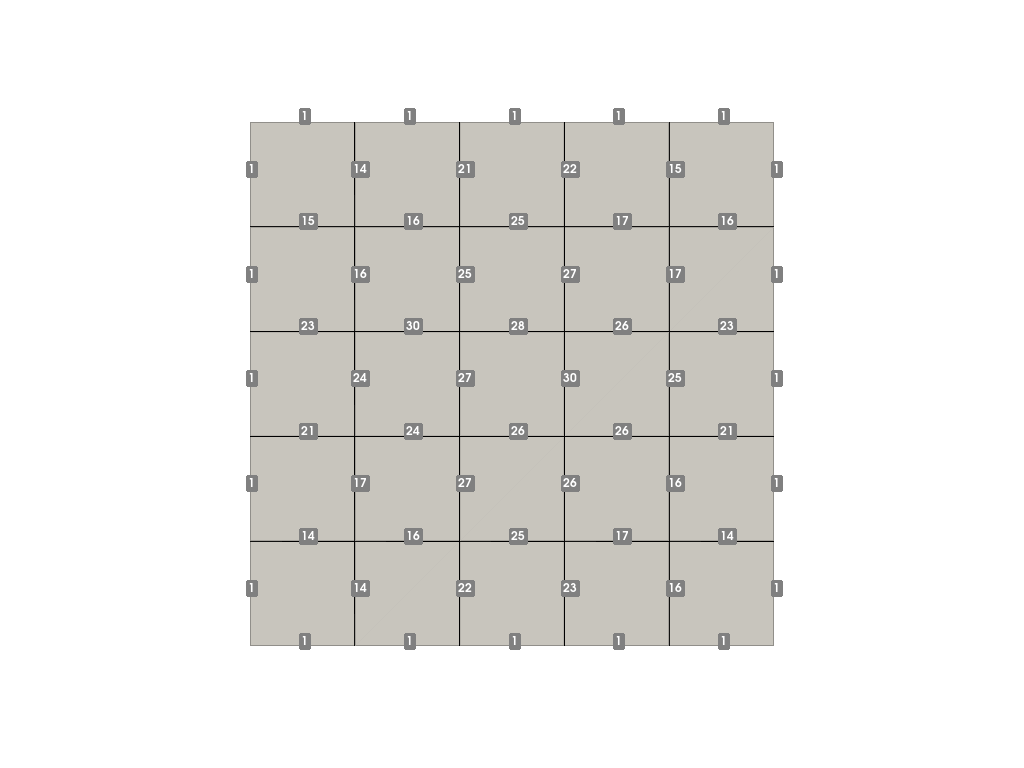}}
	\hfill
    \subfloat[{\normalsize{Correlated sampling}\label{fig:block_max_modes_mvn}}]{\includegraphics[width=0.5\linewidth]{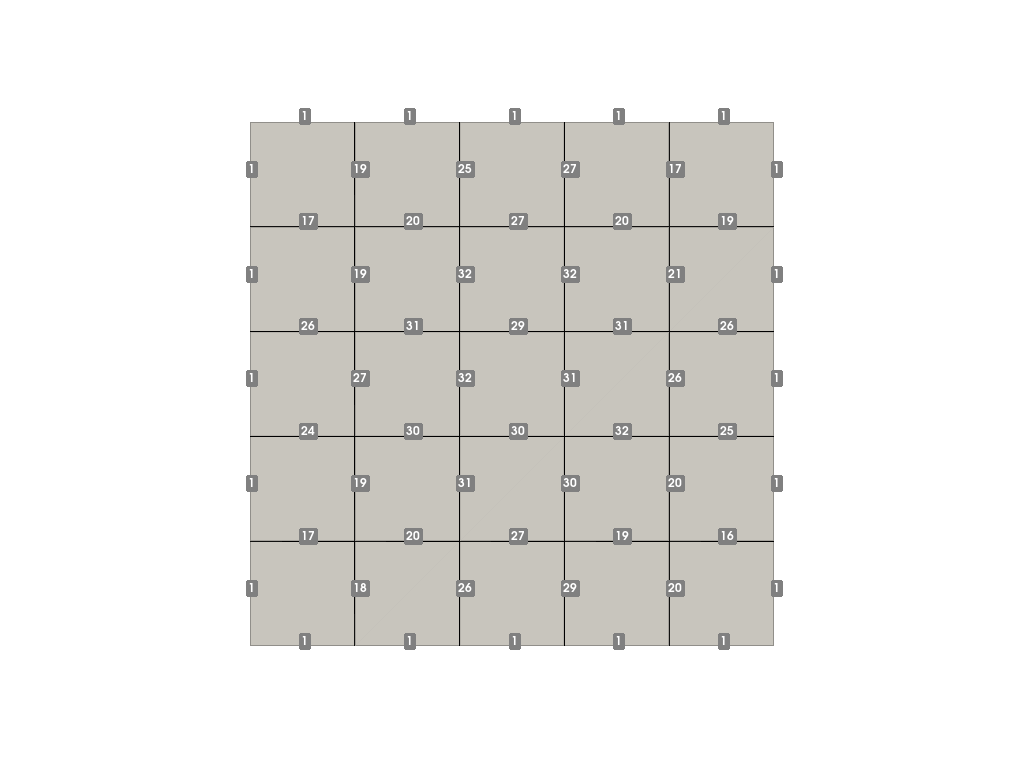}}
    \caption{Block example: number of basis functions per edge generated by~\cref{algo:rrf} for different sampling distributions and a target tolerance $\texttt{ttol}=0.001$.
    The values are given for a specific realization.}%
	\label{fig:block_num_max_modes}
\end{figure}

For one of the realizations, the number of fine scale basis functions per edge
obtained prescribing a target tolerance of $\texttt{ttol}=0.001$ is shown for both
sampling distributions
in~\cref{fig:block_num_max_modes}.
Only a single mode is necessary on the edges of the boundary of the domain,
to account for the inhomogeneous Dirichlet boundary conditions.
For both distributions, the number of fine scale basis functions in the interior
is higher than for edges close to the boundary of the domain.

\begin{figure}[htb]
	\centering
    \subfloat[\normalsize Number of DOFs and training samples against mean global relative error. The number of training samples is averaged over the number of oversampling problems. \label{fig:block_xerr}]{\includegraphics[width=0.4\linewidth]{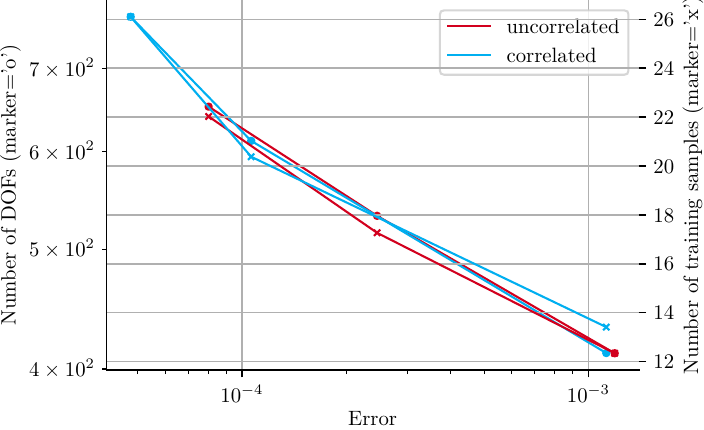}}
	\hfill
    \subfloat[\normalsize Mean global relative error and number of DOFs against target tolerance \texttt{ttol}.\label{fig:block_xttol}]{\includegraphics[width=0.4\linewidth]{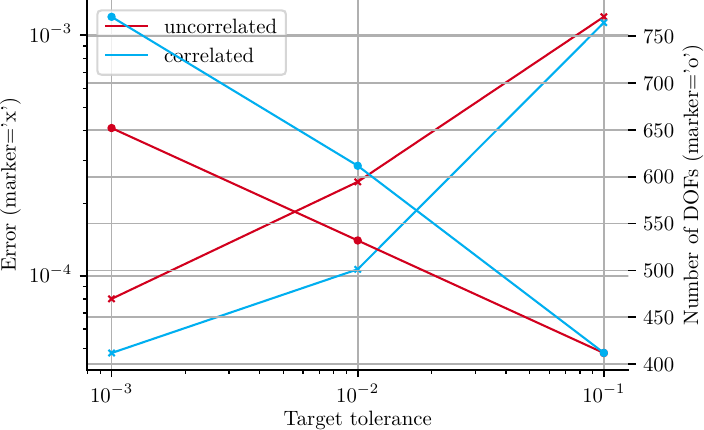}}
	\caption{Results for the block example using the maximum number of modes per edge.}%
	\label{fig:block_2y}
\end{figure}

Furthermore, using the \replaced[id=pd]{correlated samples}{\textit{multivariate normal} distribution} the number of
modes per edge that are generated is generally larger.
In this case, the criterion to exit the range finder algorithm~(\cref{algo:rrf})
is met after a larger number of training samples, which results in a more
accurate approximation, but also larger number of DoFs in the global ROM
(compare e.\,g.\, the relative error and number of DoFs for a target
tolerance of $\texttt{ttol}=0.01$ in~\cref{fig:block_2y}).
However, \cref{fig:block_xerr} shows, that to achieve a global relative error
of e.\,g.\, $1\cdot\,10^{-4}$ on average about the same number
of training samples are needed for both \replaced[id=pd]{sampling approaches}{distributions}.

\begin{figure}[htb]
    \centering
    \subfloat[{\normalsize{Uncorrelated sampling}\label{fig:block_error_fields_normal}}]{\includegraphics[width=0.3\linewidth]{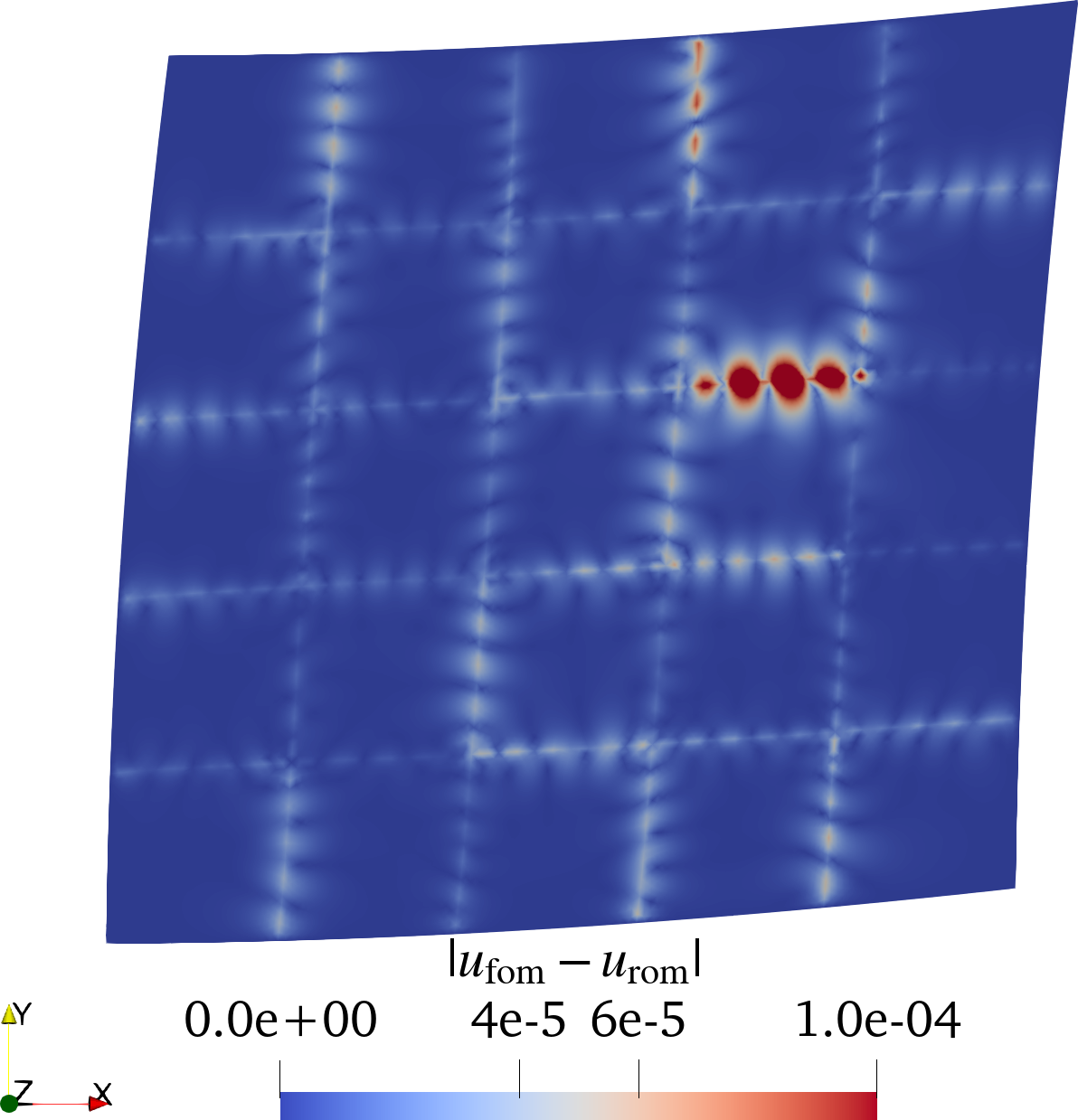}}
    \hfill
    \subfloat[{\normalsize{Correlated sampling}\label{fig:block_error_fields_multivariate_normal}}]{\includegraphics[width=0.3\linewidth]{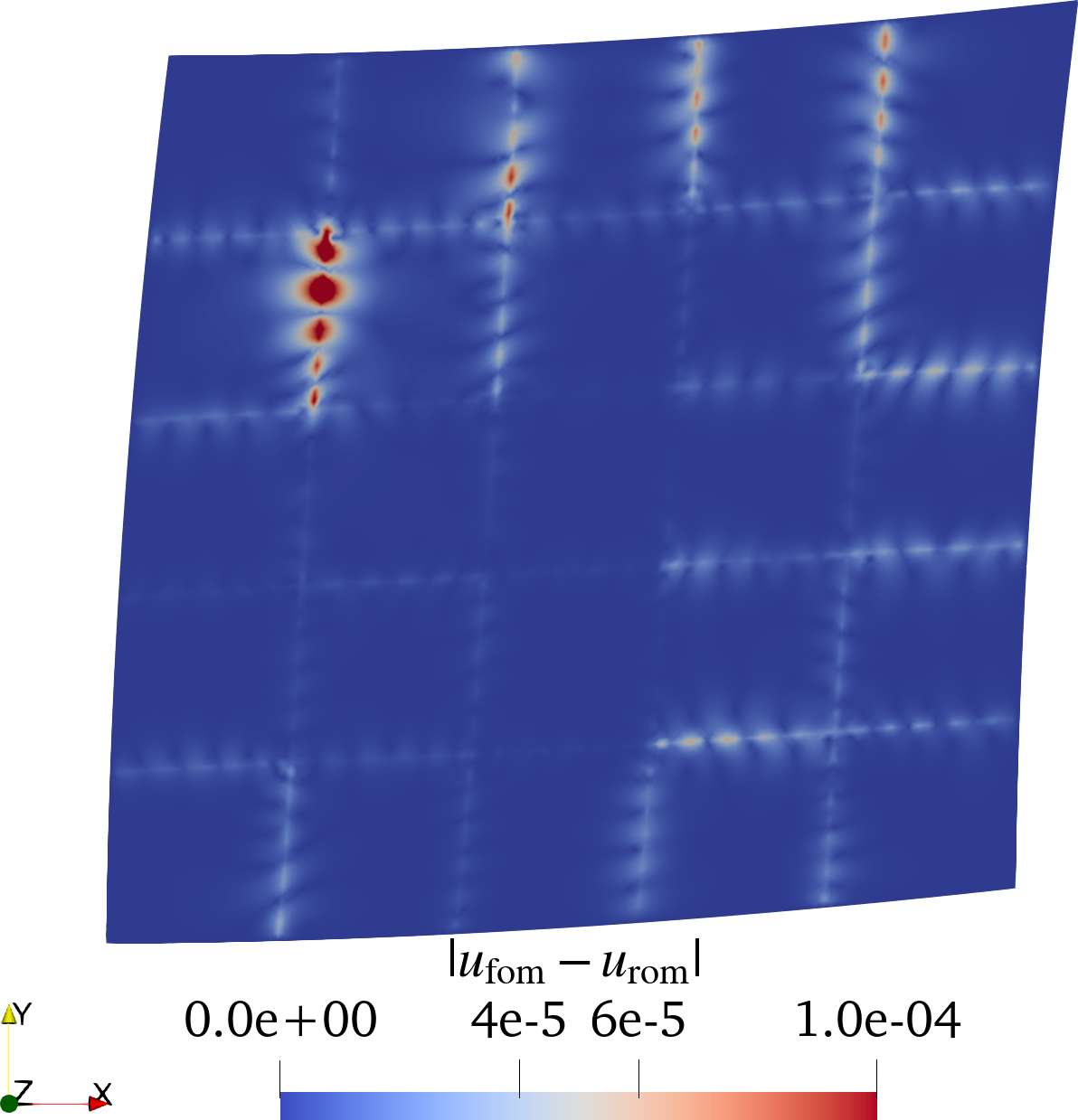}}%
    \caption{Absolute displacement error for a particular realization with different sampling distributions in the deformed placement using $10$ fine scale basis functions per edge.
    In both cases, a scale factor of $1$ is used for the deformation and the domain is scaled by the ROM displacement solution.}%
    \label{fig:block_contour}
\end{figure}

Finally, \cref{fig:block_contour} shows the absolute displacement error
for particular realizations.
For the same number of fine scale basis functions
(and in this case also number of DoFs in the global ROM),
the displacement field in the
interior is better approximated when using the \replaced[id=pd]{correlated samples}{\textit{multivariate normal}
distribution}, although the overall quality of the approximation is the same.

\FloatBarrier%
\subsection{Beam example}%
\begin{figure}[htb]
	\centering
	\subfloat[\label{fig:beam_sketch}]{\includegraphics[width=0.4\linewidth]{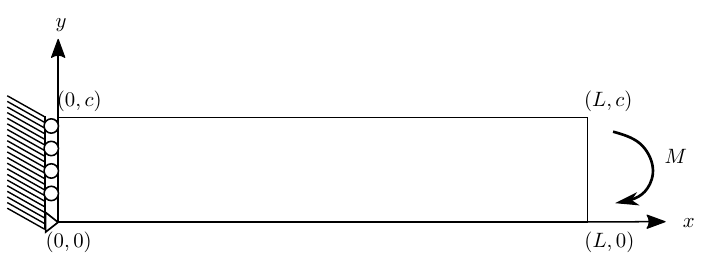}}
	\hfill
	\subfloat[\label{fig:beam_coarse_grid}]{\includegraphics[width=0.4\linewidth]{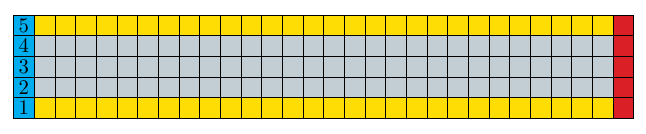}}
	\caption{Schematic representation (a) of the beam problem and coarse grid discretization showing configurations to be considered in the offline phase.}%
	\label{fig:beam_sketch_config}
\end{figure}

\label{sub:beam_example}
A beam under the state of pure bending is considered on the domain
$\varOmega_{\mathrm{gl}} = (0, L)\times(0, c)$ with length $L$, height $c$, and
thickness $t=1\,\mathrm{mm}$ as illustrated in \cref{fig:beam_sketch}.
The bending moment $M=20tc^2\,\mathrm{MPa}$ results from a horizontal
distributed force $f_x = t\left(\frac{240y}{c} - 120\right)t\,\mathrm{MPa}$, such that the analytical solution
(according to~\cite{LeeBathe1993}) in the case of a homogeneous
isotropic and linear elastic material is
\begin{align}
	\label{eq:beam_solution}
    \sigma_{xx} &= \left(\frac{240y}{c} - 120\right)\,\mathrm{MPa}, \quad \sigma_{yy} = \tau_{xy} = 0\,\mathrm{MPa},\\
    u &= \left(\frac{240}{c} xy - 120 x\right) \frac{\mathrm{MPa}}{E},\\
    v &= -\frac{\nu}{E} \left(\frac{120}{c}y^2 - 120y\right)\,\mathrm{MPa} - \frac{1}{E} \frac{120}{c}x^2\,\mathrm{MPa}\,.
\end{align}
In this example, the empirical basis (using the \replaced[id=pd]{correlated }{\textit{multivariate normal}} as
well as the \replaced[id=pd]{uncorrelated sampling approach}{\textit{normal} distribution}) is compared to the hierarchical basis
for varying ratios $\nicefrac{E_{\mathrm{a}}}{E_{\mathrm{m}}}$ of Young's moduli
of the aggregates and matrix.

The dimensions of the beam $L=1000\,\mathrm{mm}$ and $c=100\,\mathrm{mm}$ are
chosen such that the
coarse grid consists of $50\times 5$ mesoscale subdomains of type II,
see \cref{fig:rce_types_b}.
Therefore, oversampling problems which take into account the Dirichlet
and Neumann boundary conditions are considered in the offline phase
which is illustrated in~\cref{fig:beam_coarse_grid}.
For each patch (oversampling domain) that contains one of the coarse grid
cells marked blue, the associated oversampling problem needs to take into
account the homogeneous Dirichlet boundary conditions.
Analogously, for coarse grid cells marked red inhomogeneous Neumann boundary
conditions need to be considered.

\begin{figure}[tb]
	\centering
	\includegraphics[keepaspectratio]{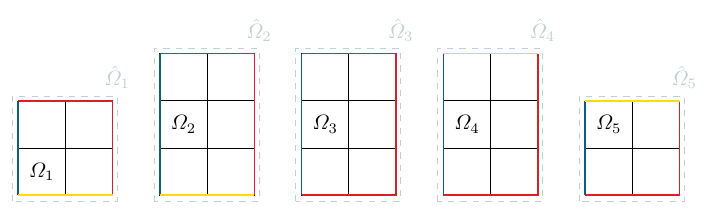}
	\caption{The configurations of the oversampling domain $\hat\varOmega$ for the subdomains $\varOmega_1, \varOmega_{2}, \varOmega_{3}$ on the left Dirichlet boundary in case of the beam example. The colors indicate the boundaries $\varSigma_{\mathrm{D}}$ (blue), $\varSigma_{\mathrm{N}}$ (yellow) and $\varGamma_{\mu}$ (red) whose topology is changing.}%
	\label{fig:beam_config_examples}
\end{figure}

The oversampling domain $\hat\varOmega_1$, $\hat\varOmega_{2}$, $\hat\varOmega_{3}$, $\hat\varOmega_{4}$ and $\hat\varOmega_{5}$, with their respective target subdomain $\varOmega_1$, $\varOmega_{2}$, $\varOmega_{3}$, $\varOmega_{4}$ and $\varOmega_{5}$ are shown in~\cref{fig:beam_config_examples} to illustrate the change in topology.

\begin{figure}[tb]
	\centering
	\includegraphics[width=0.8\linewidth]{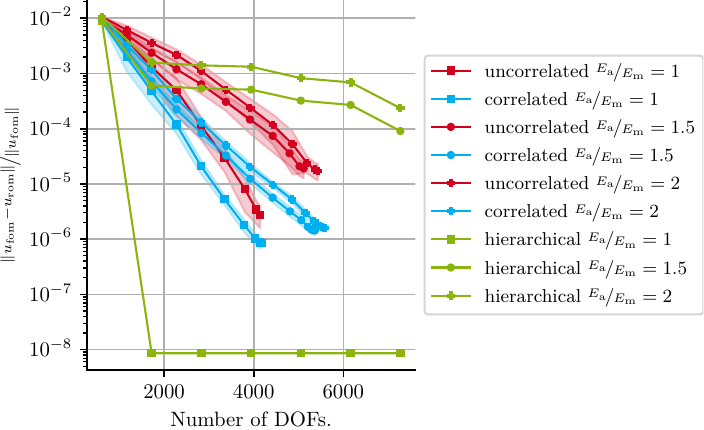}
    \caption{Beam example: \added[id=pd]{global} relative error in the
        $H^1$-norm against number of degrees of freedom.
        The empirical basis (\replaced[id=pd]{correlated and
        uncorrelated sampling approach}{\textit{multivariate normal} and \textit{normal}
        distribution} both with target tolerance $\texttt{ttol}=0.1$) and hierarchical
        basis are compared for varying ratios of Young's moduli.
        Regarding the randomized approaches, the average values over the number of \added{$\NumReal{}$}
        realizations \added[id=pd]{and the standard deviation (shaded)} of the global relative error are displayed.}%
	\label{fig:beam_rel_err}
\end{figure}

For varying ratios $\nicefrac{E_{\mathrm{a}}}{E_{\mathrm{m}}}$ of Young's moduli
for the aggregates and matrix, the global relative error is shown
in~\cref{fig:beam_rel_err}.
It is referred to~\cref{sub:block_example} for details on the calculation of
the global relative error.

In the homogeneous case, i.e. $\nicefrac{E_{\mathrm{a}}}{E_{\mathrm{m}}}=1$, the 
hierarchical basis results in a nested
finite element method and the analytical solution can be exactly represented
using only two basis functions per edge.
However, for ratios $\nicefrac{E_{\mathrm{a}}}{E_{\mathrm{m}}}>1$, the hierarchical
basis does not yield a good approximation of the fine scale part of the displacement,
whereas the relative error decays much faster in case of the empirical basis.
Moreover, in this particular example, the relative error decays at a higher
rate using the \replaced[id=pd]{correlated sampling approach}{\textit{multivariate normal} distribution} compared
to the \replaced[id=pd]{uncorrelated sampling approach}{\textit{normal} distribution}.
For the approximation of the macroscopic state of pure bending only few basis
functions are needed.
Also, as mentioned in~\cref{sub:block_example}, the algorithm tends to
generate more basis functions than would have been necessary to
achieve the target tolerance on the projection error and this
effect seems to be even stronger in case of the \replaced[id=pd]{correlated samples}{\textit{multivariate normal}
distribution}, as the relative error is almost a factor $~10$
(for $\nicefrac{E_{\mathrm{a}}}{E_{\mathrm{m}}}=2$) smaller
than the error obtained using the \replaced[id=pd]{uncorrelated samples}{\textit{normal} distribution}.

\begin{figure}[htb]
	\centering
	\includegraphics[width=0.4\linewidth]{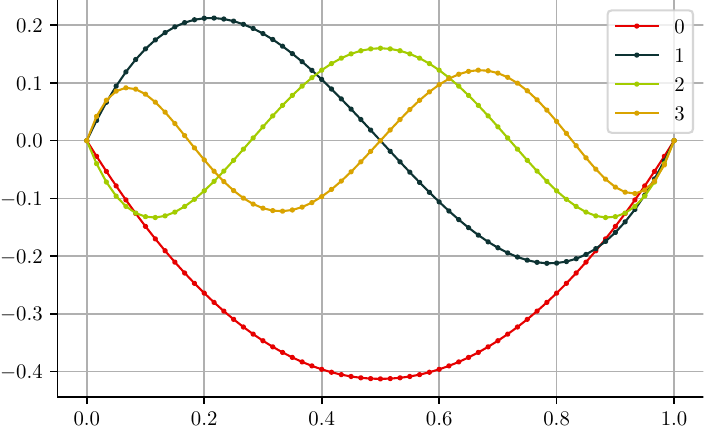}
	\caption{Illustration of the hierarchical fine scale edge basis functions.}%
	\label{fig:beam_modes_hier}
\end{figure}

\begin{figure}[htb]
	\centering
	\subfloat[\normalsize$x$-component\label{fig:modes_emp_x_1}]{\includegraphics[width=0.4\linewidth]{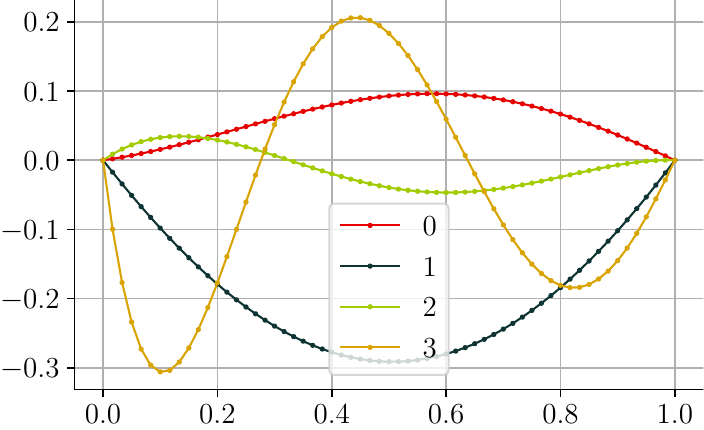}}
	\hfill
	\subfloat[\normalsize$y$-component\label{fig:modes_emp_y_1}]{\includegraphics[width=0.4\linewidth]{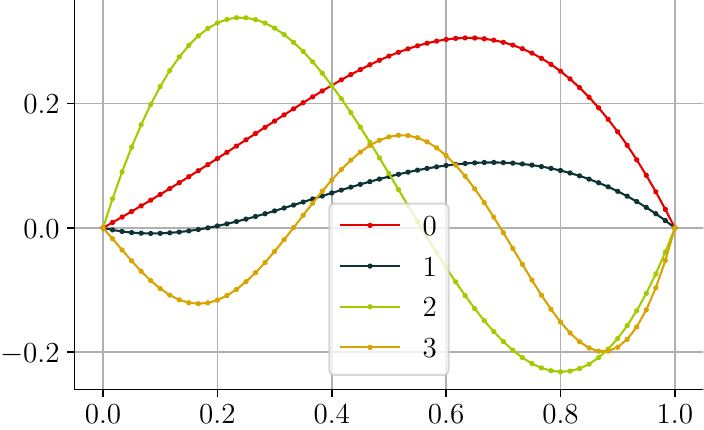}}
    \caption{\replaced[id=pd]{F}{POD f}ine scale edge basis for the empirical basis generated with correlated samples, $\texttt{ttol}=0.1$ and $\nicefrac{E_{\mathrm{a}}}{E_{\mathrm{m}}}=1$.}%
	\label{fig:beam_modes_r1}
\end{figure}

\begin{figure}[htb]
	\centering
	\subfloat[\normalsize$x$-component\label{fig:modes_emp_x}]{\includegraphics[width=0.4\linewidth]{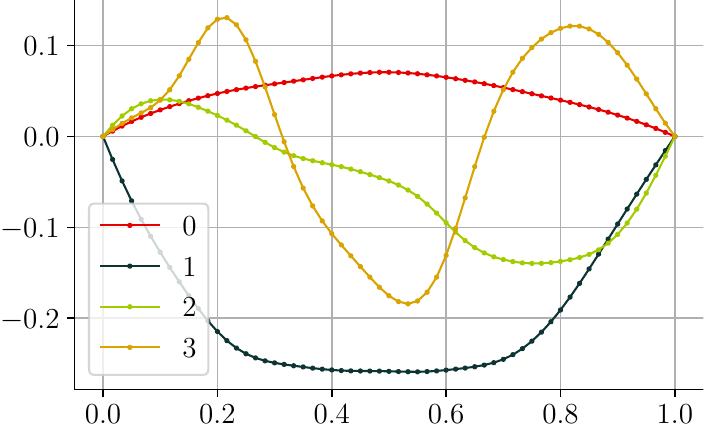}}
	\hfill
	\subfloat[\normalsize$y$-component\label{fig:modes_emp_y}]{\includegraphics[width=0.4\linewidth]{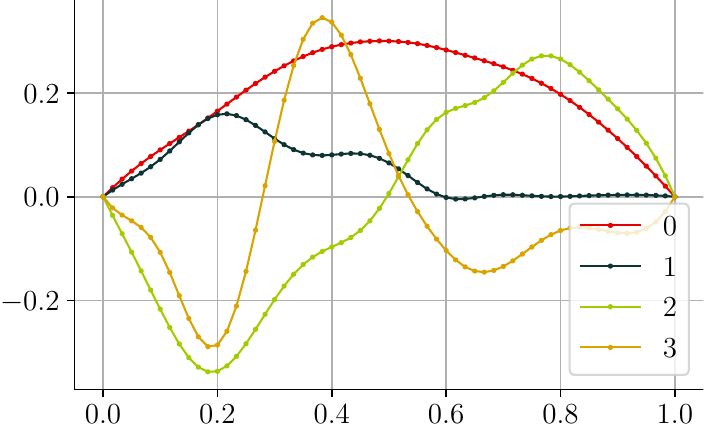}}
    \caption{\replaced[id=pd]{F}{POD f}ine scale edge basis for the empirical basis generated with correlated samples, $\texttt{ttol}=0.1$ and $\nicefrac{E_{\mathrm{a}}}{E_{\mathrm{m}}}=2$.}%
	\label{fig:beam_modes_r2}
\end{figure}

Furthermore, the purpose of this example is to illustrate the effect
of the heterogeneity on the displacement field, by comparing
the hierarchical edge basis
functions to the empirical fine scale edge basis functions.
Therefore, the hierarchical edge basis functions are plotted 
in~\cref{fig:beam_modes_hier} and the $x$- and $y$-components 
of the empirical edge basis functions
(\replaced[id=pd]{correlated sampling approach}{\textit{multivariate normal} distribution})
for a particular subdomain in the interior of the global domain
and its bottom edge
are plotted in the~\cref{fig:beam_modes_r1,fig:beam_modes_r2}.
The referenced figures show that for the homogeneous material the empirical 
fine scale edge functions share features similar to the ones of the
hierarchical shape functions.
In the heterogeneous case the overall shape of at least the first mode
is still similar, although a clear effect of the heterogeneity can be
seen, which is also stronger for the higher modes.

\FloatBarrier%
\subsection{L-panel example}%
\begin{figure}
	\centering
	\subfloat[\label{fig:lpanel_sketch}]{\includegraphics[width=0.59\linewidth]{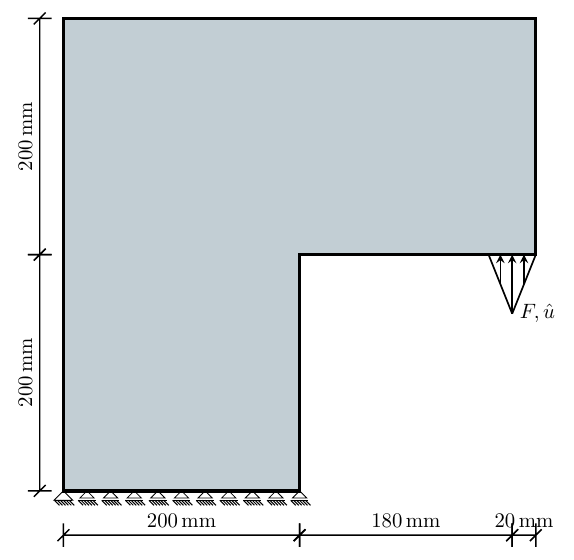}}\qquad
	\subfloat[\label{fig:lpanel_config}]{\includegraphics[width=0.59\linewidth]{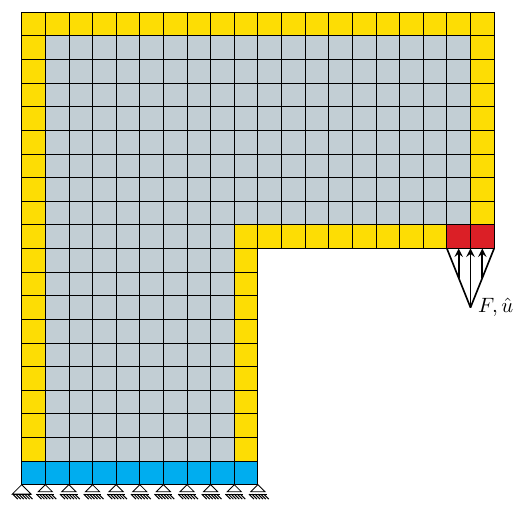}}%
	\caption{Schematic representation (a) and different configurations present in the chosen coarse grid (b) of the L-Panel example.}%
	\label{fig:lpanel_sketch_config}
\end{figure}

\begin{figure}[tb]
	\centering
	\includegraphics[width=0.8\linewidth]{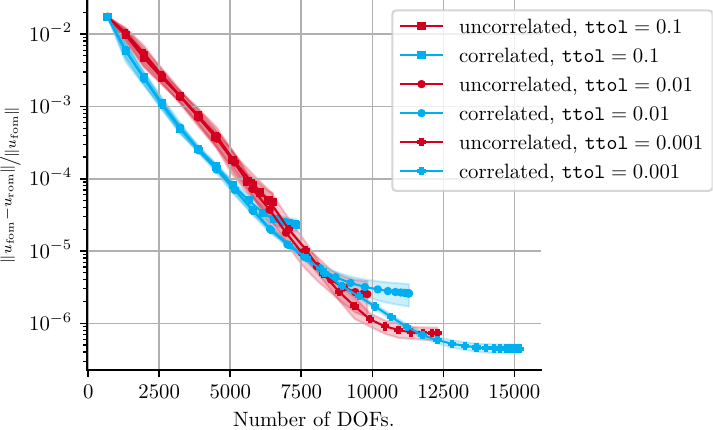}
    \caption{L-Panel example: global relative error in the $H^1$-norm
        against number of degrees of freedom in the ROM.
        \added[id=pd]{The values are averaged over the number of $\NumReal{}$ realizations.}
        \added[id=pd]{The shaded areas indicate the standard deviation of the global relative error.}
    }%
	\label{fig:lpanel_relerr}
\end{figure}

\label{sub:l_panel_example}
The third example is an L-shaped panel, see \cref{fig:lpanel_sketch}.
It features a more complex geometry and a concentrated load modeled
as a Neumann boundary condition.
For simplicity, the load is modeled as a linearly varying hat function
with maximum value $t_y=200\;\nicefrac{\mathrm{N}}{\mathrm{mm}^2}$
distributed over two coarse grid cells, such that the maximum vertical
displacement is $u_y \approx 5.4\,\mathrm{mm}$. 
Mesoscale subdomain type II (see \cref{fig:rce_types_b}) is used,
leading to the coarse grid and configurations as shown in \cref{fig:lpanel_config}.
The different configurations are encoded by color in the same manner as in the beam
example (\cref{sub:beam_example}).
In case of the L-Panel, the \replaced[id=pd]{correlated sampling approach}{\textit{multivariate normal} distribution} is
compared to the \replaced[id=pd]{uncorrelated sampling approach}{\textit{normal} distribution} for the empirical basis.
The decay of the global relative error is shown in~\cref{fig:lpanel_relerr} and
the basis generated \replaced[id=pd]{with correlated samples}{by sampling from the \textit{multivariate normal} distribution}
performs better for a smaller number of degrees of freedom
(smaller number of fine scale edge modes).
For a target tolerance of $\texttt{ttol}=0.1$ using the \replaced[id=pd]{correlated sampling approach}{\textit{multivariate normal}
distribution} more basis functions are generated, but the error of
the global approximation compared to the FOM is smaller.
In contrast, for smaller target tolerances, after reaching a certain accuracy
of the approximation, more and more basis functions are generated
for specific subdomains (edges)
due to the pessimistic error estimator of \cref{algo:rrf} without
improving the global error.
This effect is more pronounced for the \replaced[id=pd]{correlated sampling approach}{\textit{multivariate normal}
distribution} compared to the \replaced[id=pd]{uncorrelated sampling approach}{\textit{normal} distribution}.

\begin{figure}[htb]
	\centering
    \subfloat[\normalsize Number of DOFs and training samples against mean global relative error. The number of training samples is averaged over the number of oversampling problems. \label{fig:lpanel_xerr}]{\includegraphics[width=0.4\linewidth]{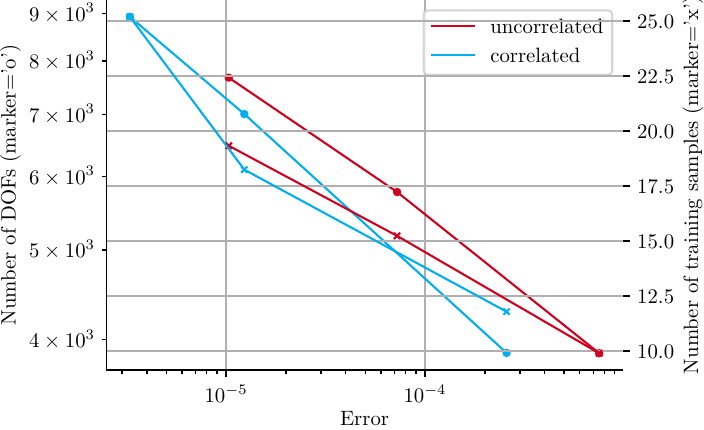}}
	\hfill
    \subfloat[\normalsize Mean global relative error and number of DOFs against target tolerance \texttt{ttol}.\label{fig:lpanel_xttol}]{\includegraphics[width=0.4\linewidth]{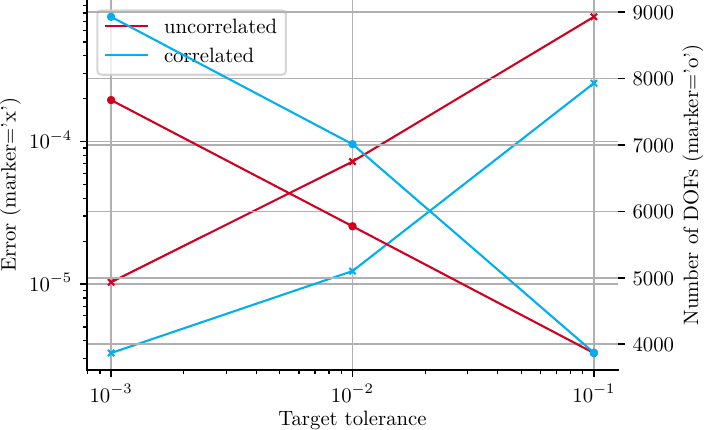}}
	\caption{Results for the L-panel example using the maximum number of modes per edge.}%
	\label{fig:lpanel_2y}
\end{figure}

\begin{figure}[htb]
    \centering
    \subfloat[\normalsize{Uncorrelated sampling}\label{fig:lpanel_error_fields_normal}]{\includegraphics[width=0.3\linewidth]{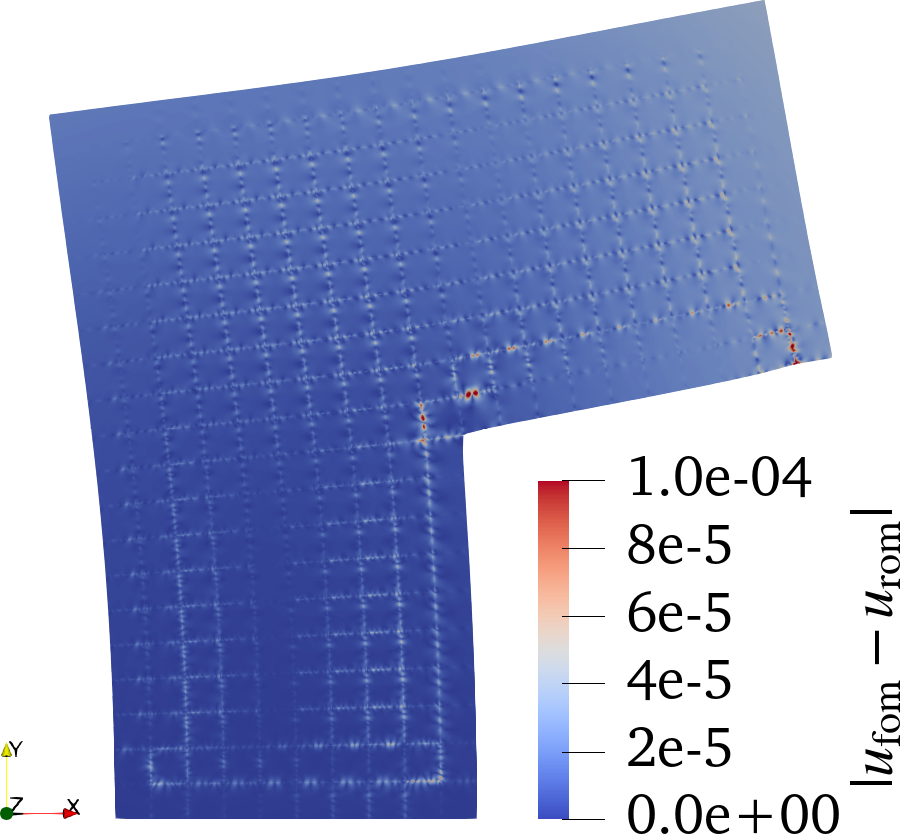}}%
    \hfill
    \subfloat[\normalsize{Correlated sampling}\label{fig:lpanel_error_fields_multivariate_normal}]{\includegraphics[width=0.3\linewidth]{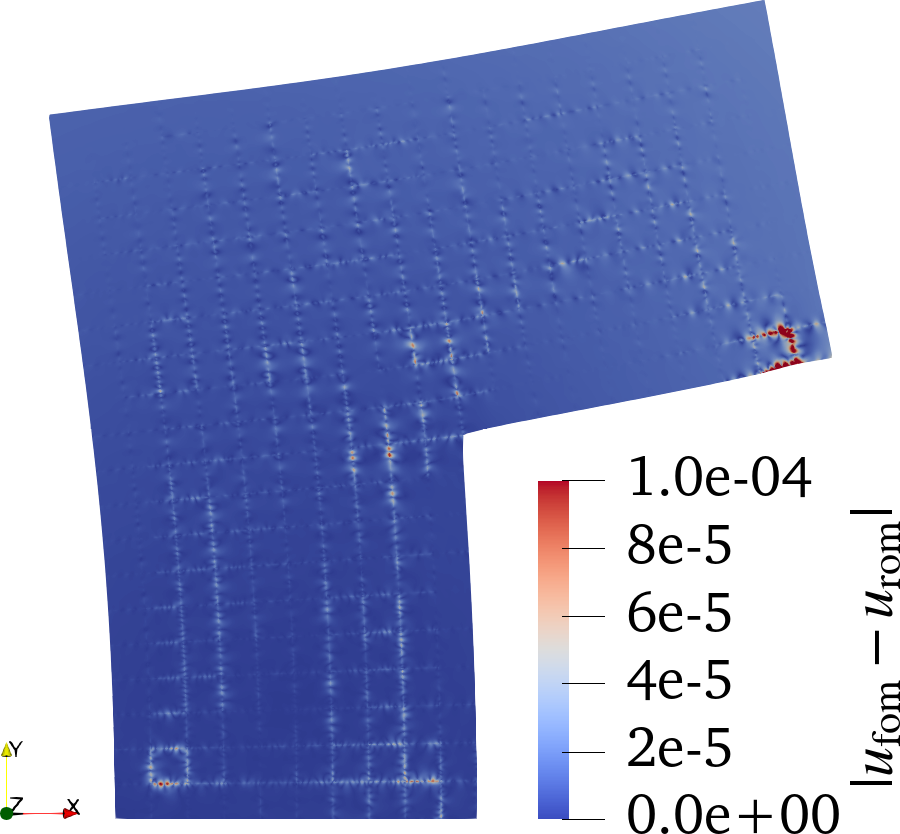}}%
    \caption{Absolute displacement error for different sampling \replaced[id=pd]{approaches}{distributions} in the deformed placement using $\LpanelFieldsNumModes{}$ fine scale basis functions per edge.
    In both cases, a scale factor of $10$ is used for the deformation and the domain is scaled by the ROM displacement solution.}%
    \label{fig:lpanel_contour}
\end{figure}

However, \cref{fig:lpanel_2y} shows the trend that 
the same approximation error can be achieved with less training samples
in the offline phase and a smaller number
of degrees of freedom in the ROM in the online phase when using
the \replaced[id=pd]{correlated sampling approach}{\textit{multivariate normal} distribution}.
Compare, e.\,g.\, the number of DoFs and training samples to achieve a global
relative error of $1\,\cdot\,10^{-4}$ (\cref{fig:lpanel_xerr}) and
the global relative error for both distributions
for a number of DoFs of (i) $~4\,000$ and (ii) $~7\,000$ (\cref{fig:lpanel_xttol}).
Moreover, \cref{fig:lpanel_contour} shows the absolute displacement error
for both sampling \replaced[id=pd]{approaches}{distributions} using a maximum of $\LpanelFieldsNumModes{}$ fine scale basis functions per edge.
While the overall level of accuracy is the same, using the \replaced[id=pd]{correlated sampling approach}{\textit{multivariate normal}
distribution} does improve the approximation near the recessed corner, which is
known to be the critical area of the structure.
On the contrary, the solution near the Neumann boundary is better approximated
using the basis constructed from uncorrelated samples\deleted[id=pd]{using the \textit{normal} distribution}.
\FloatBarrier%
\subsection{Basis construction and computational time}%
\label{sub:computational_time}

All simulations, comprising FOM and ROM (offline as well as online), are carried out using a single core (Intel Core i7-10700) and serial implementations.
It is noted that the current implementation is not optimized and the provided numbers regarding computation time serve as a proof of concept rather than a rigorous comparison of the computational performance of both models.

Regarding the offline phase, in each of the oversampling problems of all examples, a failure tolerance $\varepsilon_{\mathrm{algofail}}=1\cdot\,10^{-15}$, and $n_t=20$ test vectors were used.
Furthermore, for the norm induced by the source and range space of the transfer operator, the $L^2$ and $H^1$ inner products were used, respectively.

A comparison of the total number of DoFs in the numerical models is given 
in~\cref{tab:dofs}.
In case of the ROM, the number of DoFs and maximum number of fine scale basis
functions is averaged over the number of realizations.
For each of the investigated examples, the computational time needed to solve
the FOM and ROM are given in~\cref{tab:timings}.
With respect to the ROM, the mean over \added[id=pd]{the number of realizations}
of the assembly time and the time needed to solve the discrete system of
equations is given for a fixed number of basis functions.
Regarding the time needed to construct the empirical basis, the minimum, maximum 
and average time over all oversampling problems are averaged over the number of 
realizations as well.

It is noted that the construction of the local reduced spaces is easily
parallelizable since there is no dependency or need to communicate between
the different oversampling problems.
Hence, the total computational time to evaluate the ROM is estimated by the
time needed to construct the local basis
for the computationally most expensive oversampling problem
(offline phase) and the time to assemble and solve the discrete system
of equations for the global problem (online phase).
The time needed to precompute the global reduced solution to inform the boundary
conditions in the oversampling problem when \replaced[id=pd]{using the correlated sampling approach}{sampling from a
\textit{multivariate normal} distribution},
is negligible compared to the runtime of the randomized range finder (rrf)
algorithm.
Furthermore, in case of the \replaced[id=pd]{correlated sampling approach}{\textit{multivariate normal} distribution},
the maximum runtime of~\cref{algo:rrf} is much higher, compared to the
\replaced[id=pd]{uncorrelated sampling approach}{\textit{normal} distribution}.
This is due to the repeated computation of the eigenvalues of the covariance
matrix $\bm{\varSigma}$ as described in~\cref{ssub:training_set}.

The computational time for the online phase of the ROM (using the maximum number 
of modes per edge) and the FOM comprise the time needed to assemble and solve
the system of equations. 
In case of the block example, the ROM is not favorable,
due to the small size of the problem.
The ROM (using the \replaced[id=pd]{correlated sampling approach}{\textit{multivariate normal} distribution})
is evaluated (using $n=6$ fine scale modes) at
least $\approx 1.5$ times faster than the FOM in case
of the L-panel example and $\approx 2$ times faster in case of
the beam example.
Depending on the accuracy required for a certain application, the savings
in computational time are higher.
We also note that --- given the computational time needed to solve the FOM ---
the investigated examples are rather toy problems and that the comparison will
be more in favor of the ROM for larger problems.
\begin{table}[tb]
	\centering
	\caption{Total number of DoFs of FOM and ROM for the example problems.
        For the ROM the total number of DoFs is given for a certain number
        of fine scale edge basis functions per edge (given in paranthesis) to be able to
        compare the two sampling approaches.
        Moreover, the average of the (global) maximum number of fine scale edge basis
        functions is given.
        The values are averaged over the number of $\NumReal{}$ realizations
    for the block and L-panel example for a target tolerance of $\texttt{ttol}=0.001$
    and for the beam example for a target tolerance of $\texttt{ttol}=0.1$ and
    $\nicefrac{E_{\mathrm{a}}}{E_{\mathrm{m}}}=2$.
}%
	\label{tab:dofs}
    \begin{tabular}{lrrr}\toprule
        \multirow{2}{*}{Example problem} & FOM & \multicolumn{2}{c}{ROM} \\ \cmidrule(lr){3-4}
                                         &     & \replaced[id=pd]{uncorrelated}{\textit{normal}} & \replaced[id=pd]{correlated}{\textit{multivariate normal}} \\ \midrule
        Block               & \BlockFomNdofs{}  & \BlockNormalRomNdofs{} (10 / \BlockNormalRomMaxModes{}) & \BlockMvnRomNdofs{} (10 / \BlockMvnRomMaxModes{})\\
        Beam                & \BeamFomNdofs{}   & \BeamNormalRomNdofs{} (6 / \BeamNormalRomMaxModes{}) & \BeamMvnRomNdofs{} (6 / \BeamMvnRomMaxModes{})   \\
        L-Panel             & \LpanelFomNdofs{} & \LpanelNormalRomNdofs{} (6 / \LpanelNormalRomMaxModes{}) & \LpanelMvnRomNdofs{} (6 / \LpanelMvnRomMaxModes{})\\
        \bottomrule
    \end{tabular}
\end{table}

\begin{table}[tb]
    \centering
    \caption{Computational time in seconds for the FOM and ROM\@.
        For each sampling \replaced[id=pd]{approach}{distribution} the minimum (min), maximum (max) and average (avg)
    runtime of the offline phase for a subdomain --- consisting of the runtime of the
    randomized range finder algorithm (rrf) and the time needed to extend (ext) the \replaced[id=pd]{edge basis functions}{POD
    modes} into the respective subdomain --- are averaged over the number of \added{$10$}
    realizations.
    The results are reported
    for the block and L-panel example for a target tolerance of $\texttt{ttol}=0.001$
    and for the beam example for a target tolerance of $\texttt{ttol}=0.1$ and
    $\nicefrac{E_{\mathrm{a}}}{E_{\mathrm{m}}}=2$.
    In case of the online phase, the runtime for the assembly and solution of the
    discrete system of equations are as well averaged over the number of realizations.
    Regarding the ROM solution, the number of fine scale edge basis functions
    is chosen as $10$ for the block example, and $6$ in case of the beam and L-panel
    examples.
}%
    \label{tab:timings}
    \begin{tabular}{lccccc}\toprule
            \multirow{3}{*}{Example\,|\,Model} & \multicolumn{5}{c}{Runtime}
            \\\cmidrule{2-6}
             & \multicolumn{3}{c}{offline (rrf\,|\,ext)} & \multicolumn{2}{c}{online}
             \\\cmidrule(lr){2-4}\cmidrule(lr){5-6}
             & min & max & avg & Assembly & Solve
             \\\midrule
            Block\,|\,ROM \added{uncorrelated} & \BlockNormalRrfMin{}\,|\,\BlockNormalExtMin{} & \BlockNormalRrfMax{}\,|\,\BlockNormalExtMax{} & \BlockNormalRrfMean{}\,|\,\BlockNormalExtMean{} & \BlockNormalRomAssembly{} & \BlockNormalRomSolve{}\\
            Block\,|\,ROM \added{correlated} & \BlockMvnRrfMin{}\,|\,\BlockMvnExtMin{} & \BlockMvnRrfMax{}\,|\,\BlockMvnExtMax{} & \BlockMvnRrfMean{}\,|\,\BlockMvnExtMean{} & \BlockMvnRomAssembly{} & \BlockMvnRomSolve{}\\
            Block\,|\,{FOM} & - & - & - & \BlockFomAssembly{} & \BlockFomSolve{}\\
            Beam\,|\,ROM \added{uncorrelated} & \BeamNormalRrfMin{}\,|\,\BeamNormalExtMin{} & \BeamNormalRrfMax{}\,|\,\BeamNormalExtMax{} & \BeamNormalRrfMean{}\,|\,\BeamNormalExtMean{} & \BeamNormalRomAssembly{} & \BeamNormalRomSolve{}\\
            Beam\,|\,ROM \added{correlated} & \BeamMvnRrfMin{}\,|\,\BeamMvnExtMin{} & \BeamMvnRrfMax{}\,|\,\BeamMvnExtMax{} & \BeamMvnRrfMean{}\,|\,\BeamMvnExtMean{} & \BeamMvnRomAssembly{} & \BeamMvnRomSolve{}\\
            Beam\,|\,{FOM} & - & - & - & \BeamFomAssembly{} & \BeamFomSolve{}\\
            L-Panel\,|\,ROM \added{uncorrelated} & \LpanelNormalRrfMin{}\,|\,\LpanelNormalExtMin{} & \LpanelNormalRrfMax{}\,|\,\LpanelNormalExtMax{} & \LpanelNormalRrfMean{}\,|\,\LpanelNormalExtMean{} & \LpanelNormalRomAssembly{} & \LpanelNormalRomSolve{}\\
            L-Panel\,|\,ROM \added{correlated} & \LpanelMvnRrfMin{}\,|\,\LpanelMvnExtMin{} & \LpanelMvnRrfMax{}\,|\,\LpanelMvnExtMax{} & \LpanelMvnRrfMean{}\,|\,\LpanelMvnExtMean{} & \LpanelMvnRomAssembly{} & \LpanelMvnRomSolve{}\\
            L-panel\,|\,{FOM} & - & - & - & \LpanelFomAssembly{} & \LpanelFomSolve{}\\\bottomrule
    \end{tabular}
\end{table}
\section{Conclusions}%
\label{sec:conclusion}
In this contribution, a methodology to model linear elastic heterogeneous
structures is presented.
A method combining the variational multiscale method, domain decomposition
and model order reduction techniques is developed and applied to model
the influence of the fine scale on the coarse scale directly, addressing
multiscale problems without a clear separation of scales.
Herein, snapshots of the displacement field for local target subdomains are
computed by solving an oversampling problem with physically
informed --- by solving a global reduced problem --- \added{correlated} as well as \added{uncorrelated} random
boundary conditions.
Based on the displacement snapshots, a fine scale edge basis is constructed
\deleted{via POD} and a conforming approximation is obtained by extending the edge
functions into the interior of the respective subdomain.
This then allows for a conforming coupling of the reduced coarse grid
elements in the framework of standard finite element assembly and
hence an easy implementation.
The resulting global system of equations is sparse and reduced in size
compared to the full order model.

According to the investigated examples, one can obtain a smaller reduction
error for the same number of fine scale basis functions when
using \replaced[id=pd]{correlated samples}{a \textit{multivariate normal} distribution (correlated samples)}
instead of \replaced[id=pd]{uncorrelated samples}{a \textit{normal} distribution (uncorrelated samples)}.
The physically informed boundary conditions in the oversampling problem are shown
to improve the approximation capabilities of the reduced local spaces for
a small number of basis functions, that is larger target tolerances.
Depending on the problem, for both sampling \replaced[id=pd]{approaches}{distributions},
the pessimistic estimate of the error
in the randomized range finder algorithm
may lead to the generation of many basis functions for certain subdomains that
do not contribute to improving the global error, but result
in more accurate approximations for that part of the domain.

This effect seems to be stronger in case of the \replaced[id=pd]{correlated sampling approach}{\textit{multivariate normal}
distribution} and dependent on the size of the oversampling problem,
one has to consider the additional cost due to the repeated computation of
the eigenvalues of the covariance matrix in the correlated sampling strategy.
Nevertheless, both these issues can be overcome by \replaced[id=pd]{adapting}{improving}
the error estimate \added[id=pd]{to the multivariate normal distribution} and employing an adaptive strategy in which the
target tolerance for each local oversampling problem is chosen
based on a tolerance on the global error, which was already
done in \textit{Example 4} of~\cite{BS2018} for the GFEM.

Means to include physical states and their variation in the
training data is promising in view of the extension of the method to the
nonlinear case, which was the motivation for this project and
is also subject of future work.
In contrast to the linear case, the choice of the correct amplitudes of the
boundary data in the oversampling problem or amplitudes of the edge modes
when extending these into the respective subdomains poses a great challenge.

Also, the extension to parameterized PDEs and
development of an offline/online framework for application
in a many-query context (e.\,g.\, uncertainty quantification)
is interesting.
Especially, in applications where e.\,g.\, material parameters
or the geometry change only in small areas of the domain,
the local reduced spaces in the remaining part of the domain
might be re-used between different evaluations of the model.

\section*{Acknowledgements}
The authors gratefully acknowledge financial support by the German Research Foundation (DFG), project number 394350870, and by the European Research Council (ERC) under the European Union's Horizon 2020 research and innovation programme (ERC Grant agreement No. 818473).

\section*{Code availability}
The complete workflow, i.\,e.\, all tasks to process and postprocess the numerical experiments described in this article, are implemented using the automation tool doit~\cite{pydoit}.
The source code necessary to \textit{reproduce} the results is published together with the open source preprint~\cite{diercks2022multiscale} of this article and publicly available.
The numerical experiments are implemented with a self-written code based on the open source computing platform FEniCS~\cite{AlnaesBlechta2015a}.

\section*{Data availability}
Data will be made available on request.

\bibliography{references}%
\end{document}